# Dynamic Switching Networks: A Dynamic, Non-local, and Time-independent Approach to Emergence

A. M. Khalili[1]

[1]*School of Creative Arts and Engineering, Staffordshire University, United Kingdom*

Email*:* *a.m.khalili@outlook.com*

**Abstract** The concept of emergence is a powerful concept to explain very complex behaviour by simple underlying rules. Existing approaches of producing emergent collective behaviour have many limitations making them unable to account for the complexity we see in the real world. In this paper we propose a new dynamic, non-local, and time independent approach that uses a network-like structure to implement the laws or the rules, where the mathematical equations representing the rules are converted to a series of switching decisions carried out by the network on the particles moving in the network. The proposed approach is used to generate patterns with different types of symmetry.

## INTRODUCTION

The concept of emergence is a powerful concept to explain complex behavior by simple underlying rules. Emergence is now an active research area in field as diverse as biology, physics, and computer science.

Some birds show strong synchronization in movements. Heppner [1, 6] proposed that the synchronization in the movements could be the result of simple rules of movement followed by each bird individually. He showed through computer simulations how organized flight developed from chaotic milling.

Ants are capable of finding the shortest path from the nest to a food source [2], [3], by using a chemical substance called pheromone. While walking, ants deposit pheromone on the way, at the same time, they follow pheromone previously deposited by other ants. At the beginning, when the ants arrive to a point, where they have to decide which path to take, they choose the path randomly. By supposing that all ants have the same speed, and since one path is shorter, more ants will visit the shorter path on average, and therefore pheromone will accumulate faster on that path. After a short time, the difference in the amount of pheromone on the two paths is large, and it will be enough to influence the decision process of new ants. Many algorithms have been inspired from this behaviour to solve many problems such as the traveling salesman problem [4], [5].

Cellular Automata (CA) are well known computational tools to study emergent collective behaviour. Cellular Automata have been used for the simulation of complex phenomena, such as growth, evolution, etc. CA were proposed by John von Neumann [7] to deal with the issue of reproduction of natural processes.

Conway [8] created the Game of Life, where he designed a two-dimensional cellular automaton with such rules to avoid the formation of structures that quickly disappear or grow freely, interesting complex behaviors have been observed such as the gliders, which are small groups of cells that appear to move like independent emergent entities. The simple rules Conway used can be summarized as follow:

- Living cells die if they have fewer than two neighbors (loneliness)
- Living cells die if they have more than three neighbors (overpopulation).
- Dead cells become alive if they have three neighbors (reproduction).
- Otherwise there are no change.

Wolfram [9] used a cellular automaton with simple rules and simple initial conditions to produce behaviours that are highly complex. Then by observing the behaviour of different rules in many simulations, he was able to classify the behaviour of the rules as: stable structures or simple periodic patterns, chaotic non-periodic behaviour, and complex patterns.

In physics, two-dimensional CA were created to study statistical properties of gases [10]. CA have also been used for simulation of fluids, or granular substances.

Konrad Zuse suggested that the universe is the output of a computation on a giant cellular automaton [11]. A growing number of physicists are now taking the view that information is the most fundamental thing [12], [13], an idea that can be traced back to John Wheeler who famously said "it from bit" [14]. In this view, all interactions between physical systems are information processing. Marcopolo et al. [15], [16], proposed a study of quantum gravity based on spin systems as toy models for emergent geometry and gravity. These models are based on quantum networks with no a priori geometric notions. Similar line of work includes [17]-[24].

Existing approaches of producing complex behaviour from simple rules have many limitations [25], [26] making them unable to account for the complexity we see in the real world. Some examples of complex phenomena we see in the real world include:

- Current events depending on non-local events, such as quantum entanglement [27], [28]. Quantum entanglement is a phenomenon that occurs when two particles are generated in a way such that the states of the particles can't be described independently of each other. One particle of an entangled pair knows instantly the result of the measurement that has been performed on the other particle, even though there is no known means for this information to be communicated between the two particles, which may be separated by a very large distance.
- Current events depending on events from the future, such as the delayed choice quantum eraser experiment [29]. In the classic double-slit experiment, no interference is observed when a measurement is performed to know which slit the photon went through, while an interference pattern is observed when no measurement is performed. However, what makes the delayed choice quantum eraser experiment different is that the choice of whether to measure or not to measure the photon 2 (which is entangled with photon 1) was not made until 8 ns after the position of photon 1 had been already recorded. Which may suggest that the knowledge of the future fate of photon 2 would determine the activity of photon 1 in its present.

These complex quantum phenomena might result from simple underlying rules, many physicists [17]-[20] have argued that there is a more fundamental theory underlying quantum mechanics and quantum mechanics might emerge from the laws of this theory.

The main contribution of this paper is to propose a new computational tool that may help in modelling more complex phenomena by allowing non-local and time-independent events to take place. The proposed approach may be more suitable to model physical systems, however it may also help in modelling complex systems where a centralized approach is required. The proposed approach can be seen as a dynamic, non-local, and time independent extension to cellular automata, it uses a network-like structure to implement the laws or the rules, where the mathematical equations representing the rules are converted to a series of switching decisions carried out by the network on the particles moving in the network. Using moving particles is more realistic in modelling physical systems where the collective behaviour emerges from the movements and the interactions of these particles.

One example is fluid dynamics where the macroscopic behaviour of the fluid is governed by rules governing the interactions of the molecules. Another example is the temperature where the rules governing the interactions of gas molecules give rise to emergent property such as the temperature. Other examples include the quantum phenomena described earlier, however modelling these phenomena is beyond the scope of this paper. In order to have a macro view on all the interactions that are taking place and to allow non-local and time independent events, a network-like entity should be in place to cover the whole space and to direct the particles movements, because the particles don't have access to the state of other particles far away from them or to the state of other particles in the future. The proposed approach is used to generate patterns with different types of symmetry.

Transfer entropy [30] has been recently used to quantify the directional flow of information in complex systems. It is shown to be useful particularly in describing distributed information processing. Transfer entropy was introduced to address limitations in the other measure such as mutual information, which is a symmetric measure of the shared information. Transfer entropy has been recently used to describe waves of motion in swarms and flocks [31]. It has been used to evolve a modular robot to maximize the information transfer between the components [32]. It has been also used to show that regular networks are usually linked with information storage, random networks are associated with information transfer, and small-world networks have a balance between the two [33]. Transfer entropy is an interesting measure that could be used to study or to optimize the information flow in the proposed framework.

Liu et al. [34] proposed an analytical tool to study the controllability of complex systems and to identify the driver nodes that can control the dynamics of the entire system. The number of driver nodes can be determined by the degree distribution of the network. Sparse inhomogeneous networks, which appear in many real complex systems, are found to be the most difficult to control. On the other hand, dense homogeneous networks could be controlled using a few driver nodes. However, since the particles could move in any place in the proposed framework and since the rules should be applied everywhere in the network, all the nodes of the network should have control and switching ability.

## Method

In this paper, we will propose a new approach in which the rules or the laws are applied. The rules will be applied on a set of particles moving in the network, instead of applying the rules on static cells. The network has two functionalities, the first one is to provide the space in which the particles move, and the second one is an active role, where the network changes the movement of the particles to satisfy the rules by sending the particle through one out of eight directions. The proposed approach goes beyond existing approaches in the following main points:

- A dynamic approach: unlike other approaches where the basic building blocks are static cells, the basic building blocks of the proposed approach are particles moving in a network-like structure, which will allow us to represent natural phenomena in a more realistic way. The laws or the rules will be implemented by changing the movement of the particles to satisfy the rules.
- A non-local approach: unlike other approaches where the cell is only affected by its adjacent cells, the proposed approach will allow the particles to be affected by particles far from their current positions.
- A time-independent approach: unlike other approaches where current events are only dependent on past events, the proposed approach will allow current events to depend on events from the future.
- An external entity approach: unlike other approaches where each cell applies the rules on its own when it interacts with other cells, in the proposed approach the rules are applied by the network, which is an external entity that has access to non-local and future events.
- A combined Micro and Macro approach: unlike other approaches where the cells are not aware of what is going on at the Macro level, the network is fully aware of both the Micro and the Macro levels.

A number of particles is assumed to be moving in straight lines inside a regular network structure, the initial positions and directions of the particles are generated using a pseudo-random number generator equation. At each node the particle should take one direction out of eight possible directions as shown in Fig. 1.

When the particles reach the boundary, they will bounce off to the interior. Unlike other approaches, events happening in the future are assumed to be known by knowing the initial positions and the equations governing the particles movement. The network will be responsible of implementing the laws by converting the mathematical equations representing the laws to a series of switching decisions. The network will be also responsible of controlling the speed of the particles. As illustrated in Fig. 2, the moving particles are assumed to be moving in a network-like structure, each particle will be switched to one out of eight directions depending on the rules. For example, if the rule dictates that two particles should repel each other when they are in adjacent nodes, the network will switch the two particles in opposite directions.

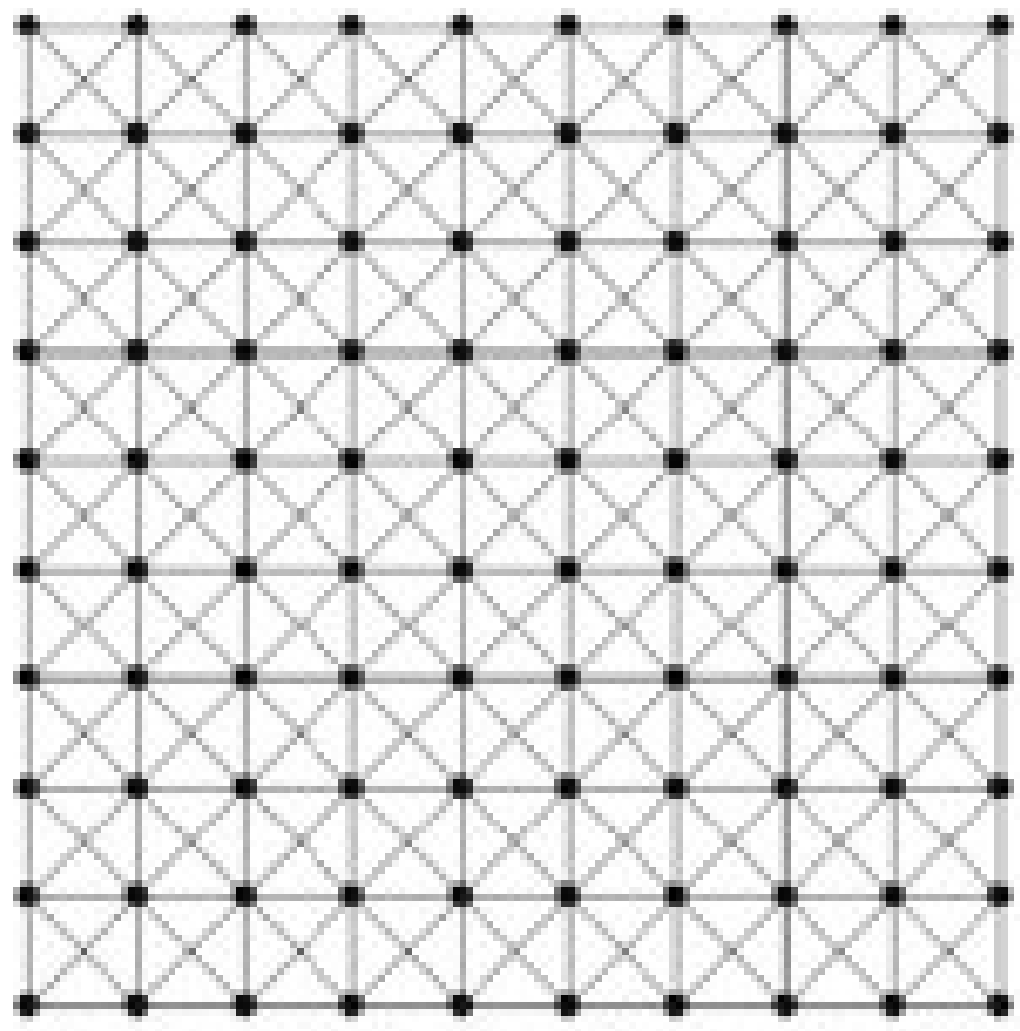

**Fig. 1.** The structure of the network.

The rules could vary from local rules governing the particles interactions such as repulsion or attraction of close particles, to global rules such as correlation between far distance particles. If no rules found to be applied on the particle, the particle will continue to move according to its own movement equation. Each node of the network has the entire code that governs the movement of all particles and their initial positions, which means that the switching decision won't be dependent only on local events such as particles interactions, but on the entire dynamics. This will enable the network to make global decisions based on the global picture instead of making decisions based only on local interactions. This will also enable the network to make decisions based on future events instead of making decisions based on past and current events. Each node is assumed to have basic information processing capability, where the node needs to retrieve the information of the particle from an information storage and then uses this information to calculate to which direction the particle should be switched. To prevent any loops or mutual dependence between current and future events, the network will prevent any current event to depend on future events if this dependence will create a loop.

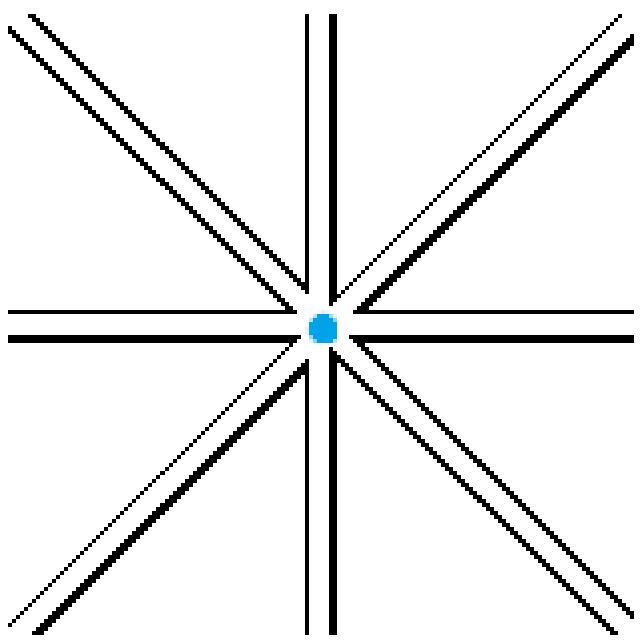

**Fig. 2.** The rules will be applied by a simple switching decision to one out of eight directions.

The best analogy to the proposed approach would be a data network, where the router at each node will route the packet through specific links to achieve some global criteria; for instance, minimising the congestion or minimising the length of the total path.

To make the approach more clear, we will provide a basic mathematical formulation, we will assume that the mathematical equation representing the law can be described by the function $f(\mathrm{t})$, where $f(\mathrm{t})$ can be defined as a series of positions $pos_t$ at different time instances t as described by (1)

$$f(1) = pos_1, f(2) = pos_2, \dots, f(\mathrm{t}) = pos_t \quad (1)$$

Similarly, the network will produce a series of switching decisions which can be expressed as a series of positions S(t), S(t) is described by (2)

$$S(1) = pos_1, S(2) = pos_2, \ldots, S(t) = pos_t \quad (2)$$

The goal of the network is to produce a series of switching decisions that minimizes the distance between $f$ and S at each time instance as described by (3). The lattice shown in Fig.1 was chosen because it is more accurate in approximating the laws than the typical four directions lattice. The optimal lattice would be the lattice that minimizes the equation given by (3).

$$\min \sum_{t} Distance(S(t), f(t)) \quad (3)$$

Where Distance is the Euclidian distance. To make the mathematical formulation more clear, let us suppose that the law dictates that the particle should follow a mathematical equation that can be represented by a series of positions $f$, these positions do not necessarily match with the positions of the nodes of the network, $f$ is shown as an orange line in Fig. 3. Then the network should produce a series of switching decisions represented by a series of positions S that best approximates $f$. The yellow points in Fig. 3 show the series of positions S that minimize the sum of the Euclidian distances between $f$ and S.

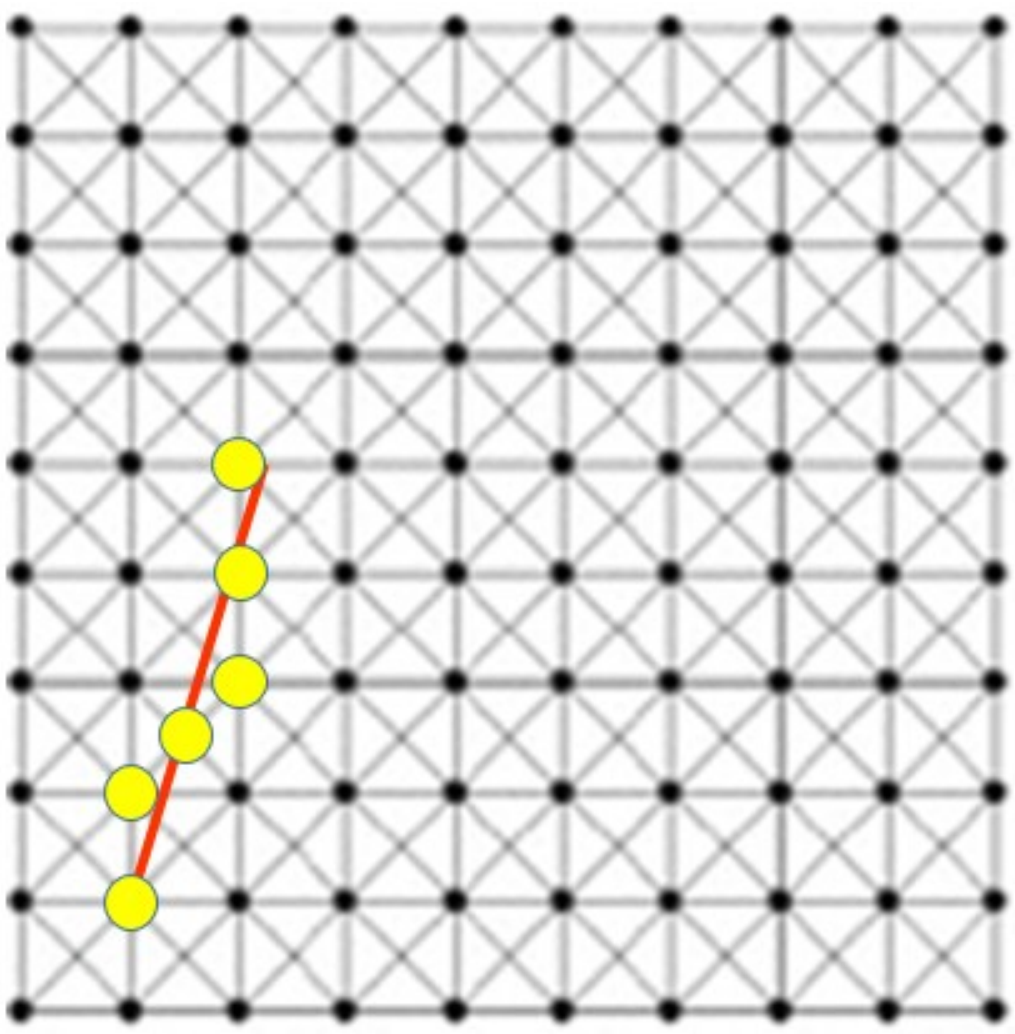

**Fig. 3.** The series of positions resulted from the rule (shown in orange) versus the series of positions resulted from the switching decisions (shown in yellow).

The role of the network could be extended to other information processing functionalities, such as: different types of particles may get different rules, or using probabilistic rules, where the rules are only applied for specific percentage of the time.

To illustrate the proposed idea further, we will take a simple example, we will assume that the goal of the network is to form symmetrical patterns from the moving particles. This example requires that the network has a global knowledge of the movement of all particles. The network will fix the positions of two particles if they have symmetry at the current time instant or at any time instance in the future. For example, the network will fix the particle $p_1$ in its position if there is any particle symmetrical to it at the current time instance or at any time instance in the future, and an identical copy of the fixed particle will continue to move using the same movement equation.

Two types of symmetry will be tested, the first one is the symmetry around the y axis, i.e.

$$x_{p2} = -x_{p1}$$

$$y_{p2} = y_{p1}$$

Where $p_1$ is the first particle and $p_2$ is the second particle. The second type is the symmetry around the y axis, the x axis, and the diagonal, i.e.

$$x_{p2} = -x_{p1}$$

$$y_{p2} = y_{p1}$$

$$x_{p3} = x_{p1}$$

$$y_{p3} = -y_{p1}$$

$$x_{p4} = -x_{p1}$$

$$y_{p4} = -y_{p1}$$

Where $p_1$ is the first particle, $p_2$ is the second particle, $p_3$ is the third particle, and $p_4$ is the fourth particle. Using the mathematical formulation described earlier, we will get $f$ = S = current position, because we make the particle stops if the symmetry rules are satisfied. In the next section we will describe the configuration and the parameters of the simulations and will show the resulted patterns.

## Simulation Results

To test the proposed approach, a number of moving particles inside a closed circle is considered. The particles will bounce off to the interior when they reach the boundary of the circle. Four parameters will be varied, the first one is the time of the movement of the particles T, the longer the time the more likely that symmetrical patterns will emerge. The second parameter is the size of the particles S, which is related to the number of particles, the larger the number of particles, the smaller the particles size. The third parameter is the number of the particles N, the higher the number the more likely that symmetrical patterns will emerge. And the fourth parameter is the scale of the network D, i.e. the number of network nodes per distance unit, the higher the number of nodes per distance unit, the less likely that symmetrical patterns will emerge. The values of these parameters are chosen such that the density of the particles of the resulted pattern is not too high or too low, so a meaningful pattern (non-random pattern) can emerge.

Multiple simulations with different parameters and different initial positions of the particles are performed. Fig.4 to Fig.13 show some of the resulted patterns when the first type of symmetry is considered, and Fig.14 to Fig.17 show some of the resulted patterns when the second type of symmetry is considered. The values of the four parameters are listed below each figure. For example, Fig.4, Fig.5, and Fig.17 show how much a particle is symmetrical with its future positions. Fig.10, and Fig. 12 show how simple symmetry rules may give rise to a face-like structure.

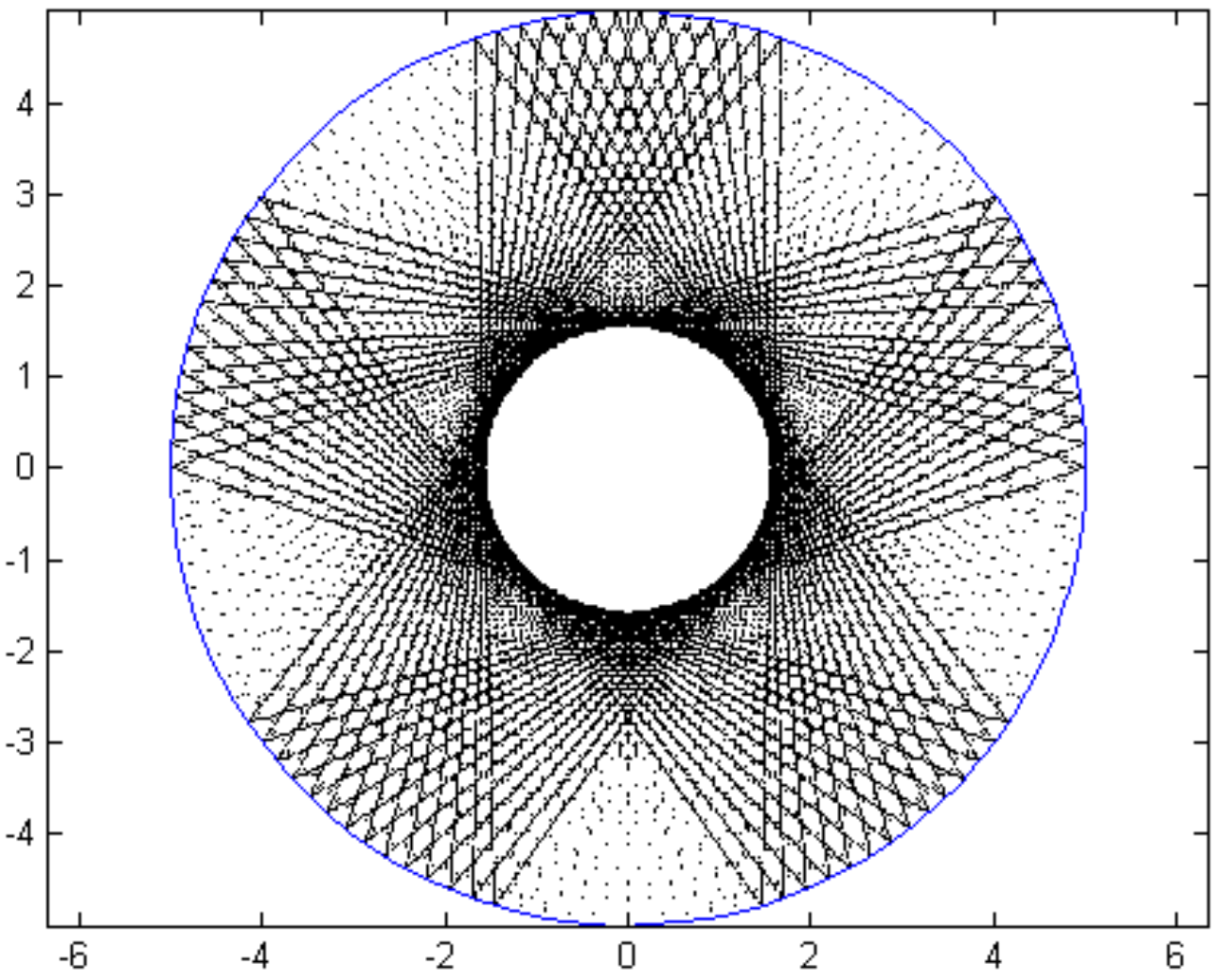

**Fig. 4.** The resulted pattern when T=300, N=1, S=3, D=50.

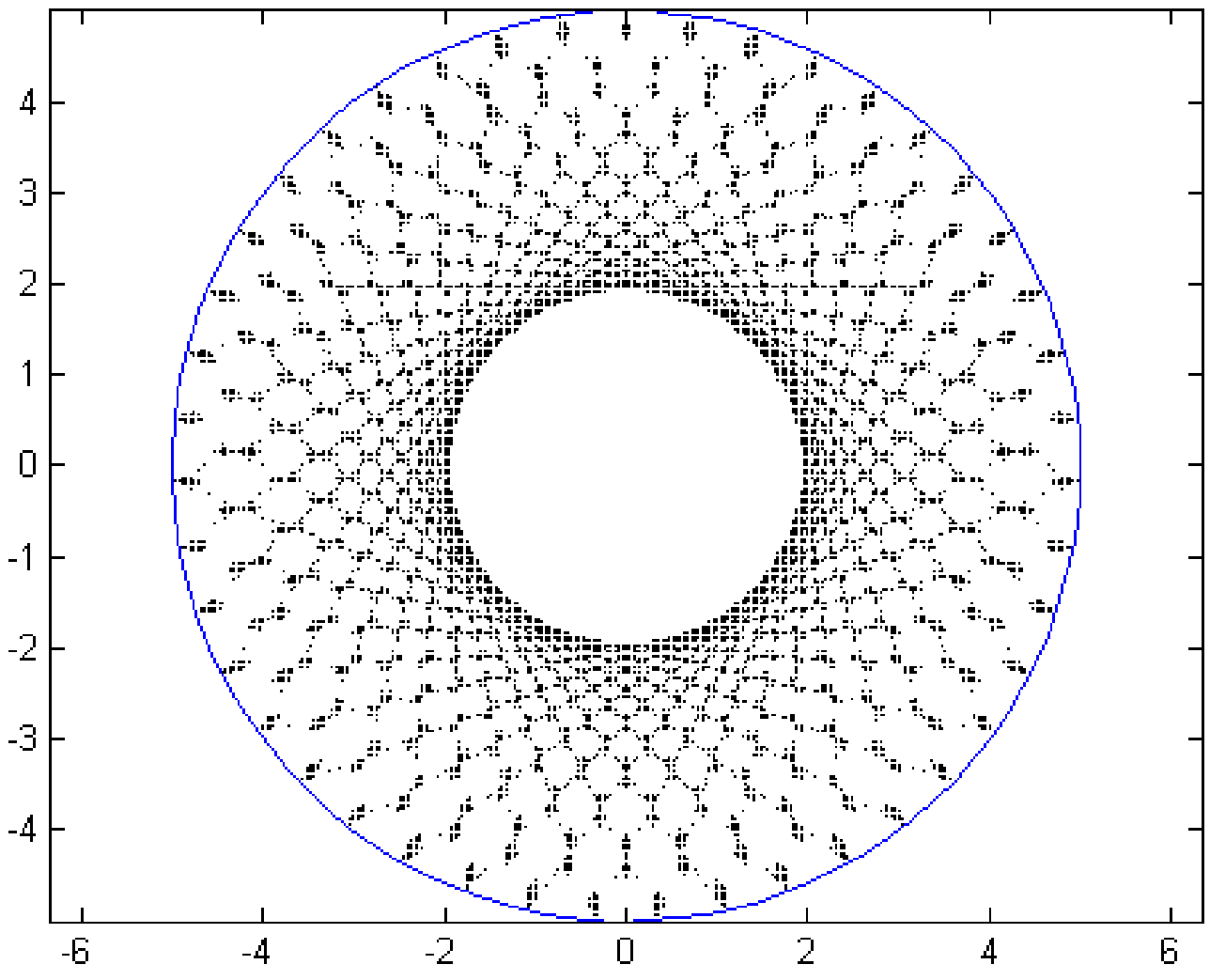

**Fig. 5.** The resulted pattern when T=300, N=1, S=3, D=100.

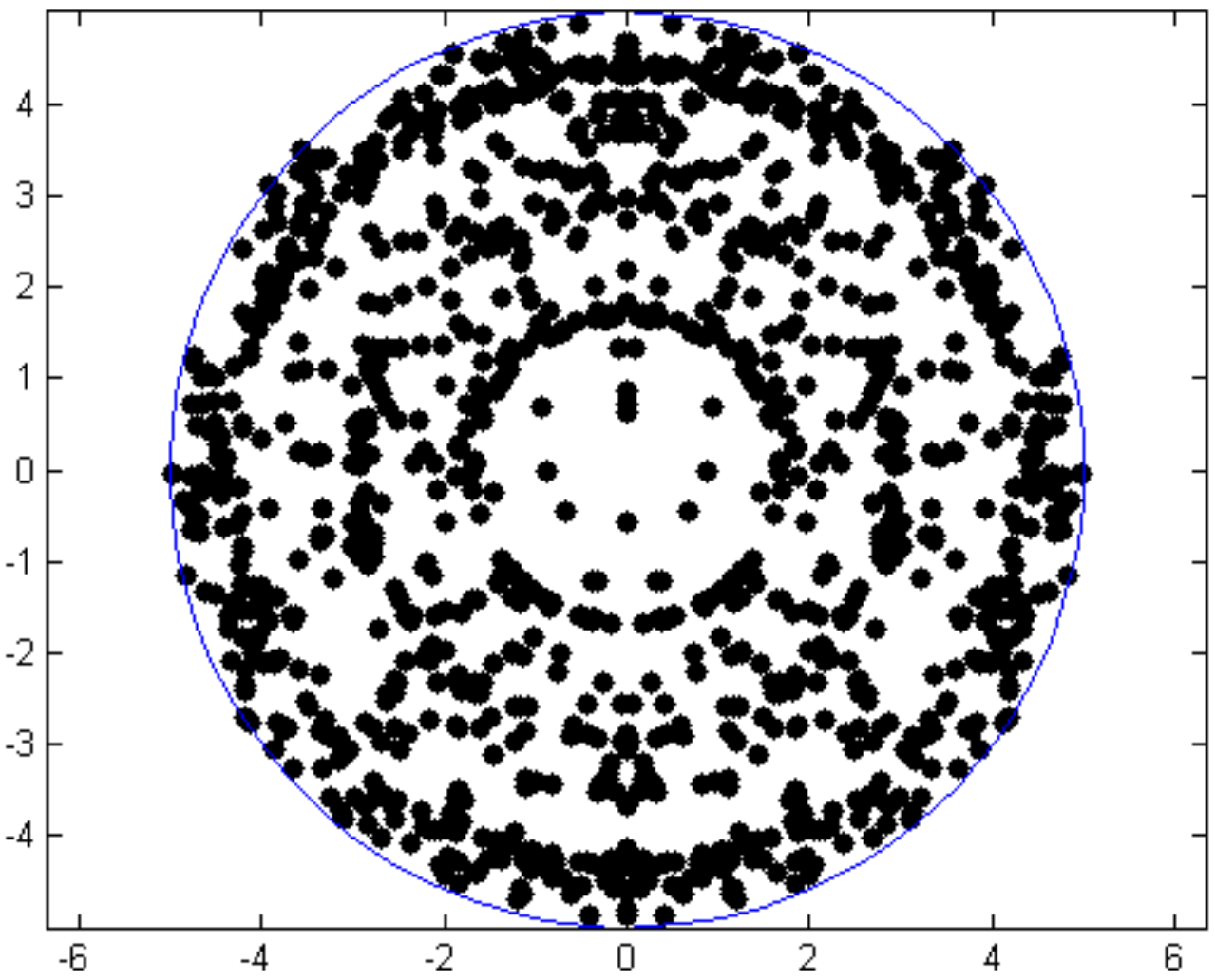

**Fig. 6.** The resulted pattern when T=30, N=20, S=20, D=600.

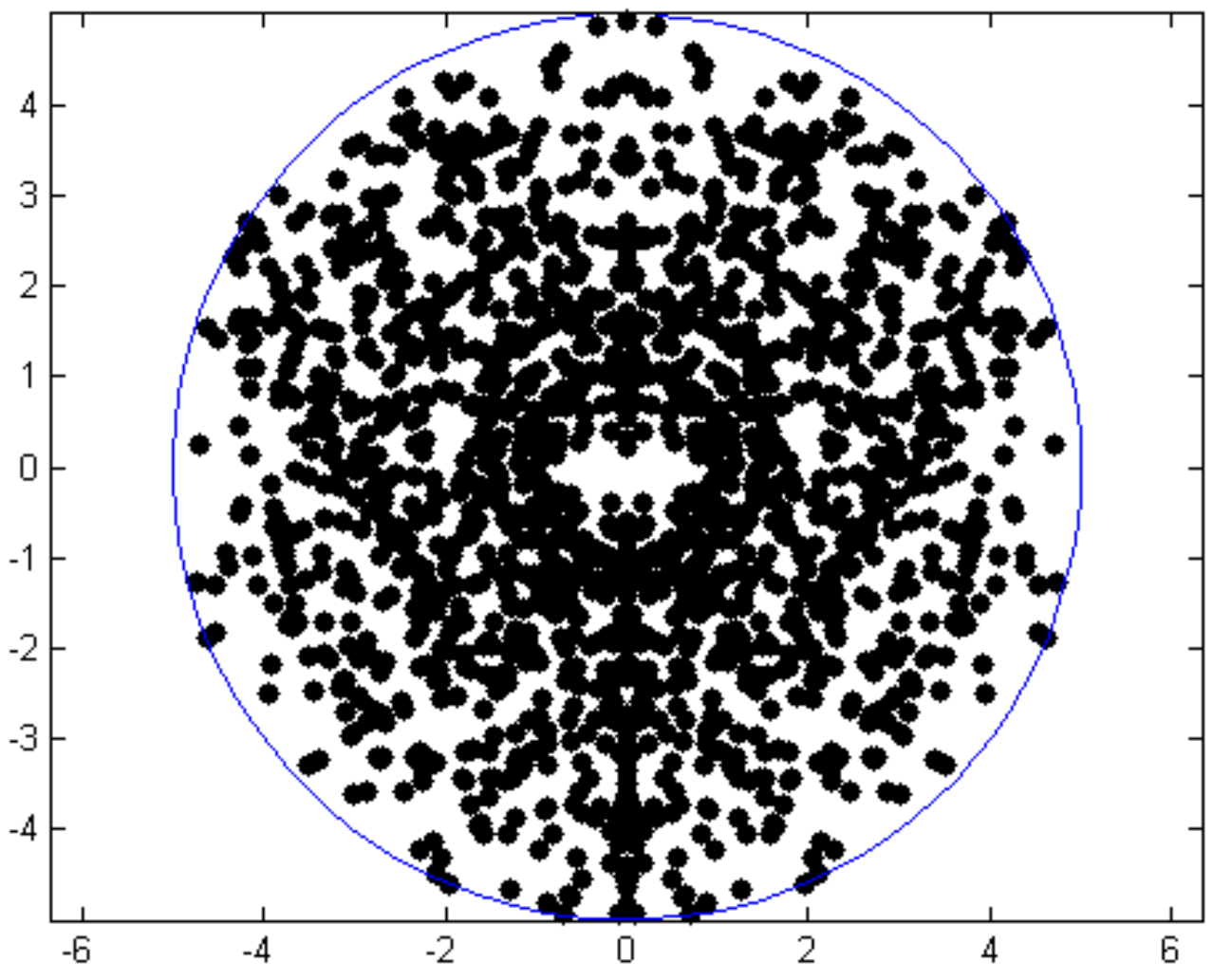

**Fig. 7.** The resulted pattern when T=30, N=20, S=20, D=400.

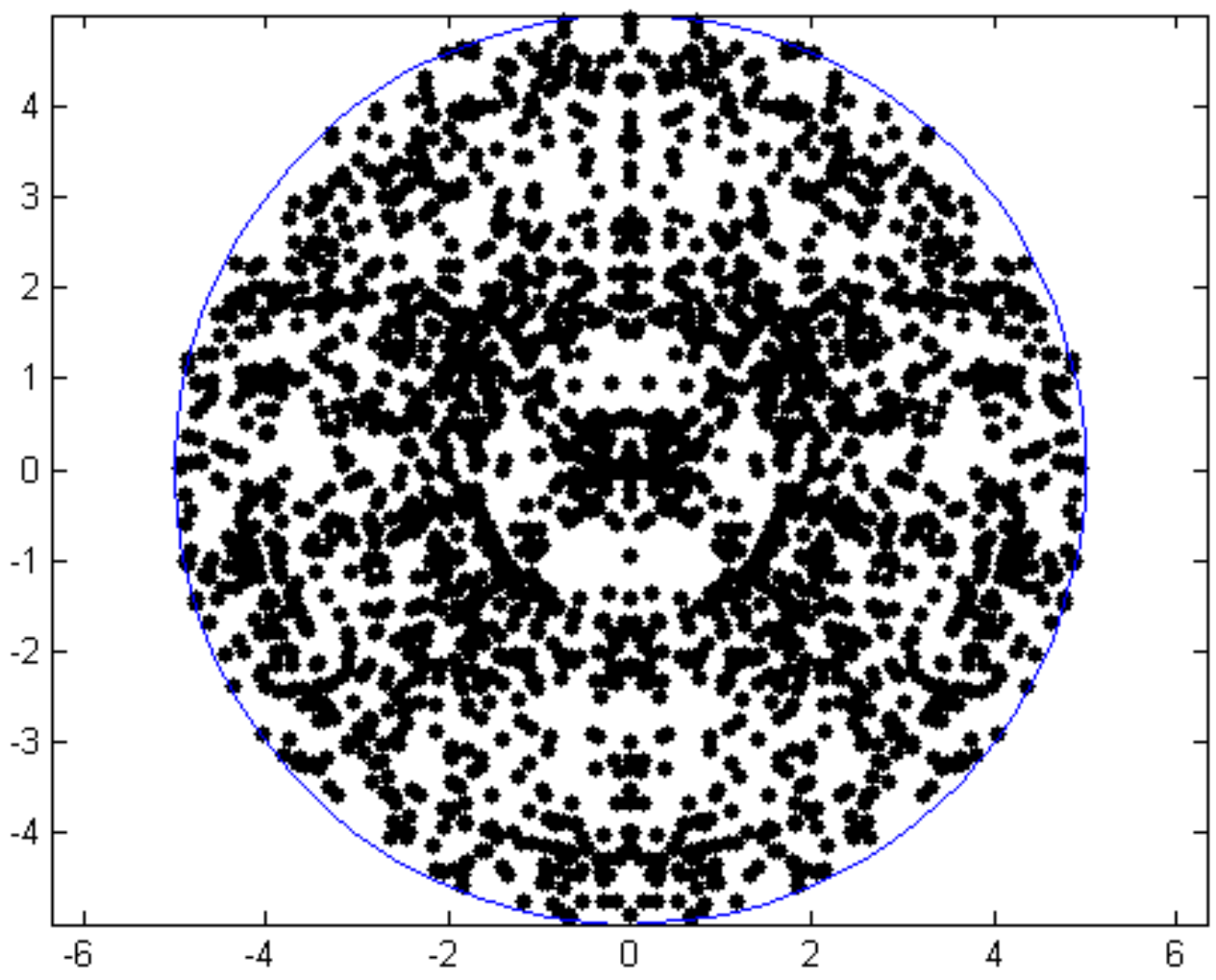

**Fig. 8.** The resulted pattern when T=30, N=20, S=15, D=350.

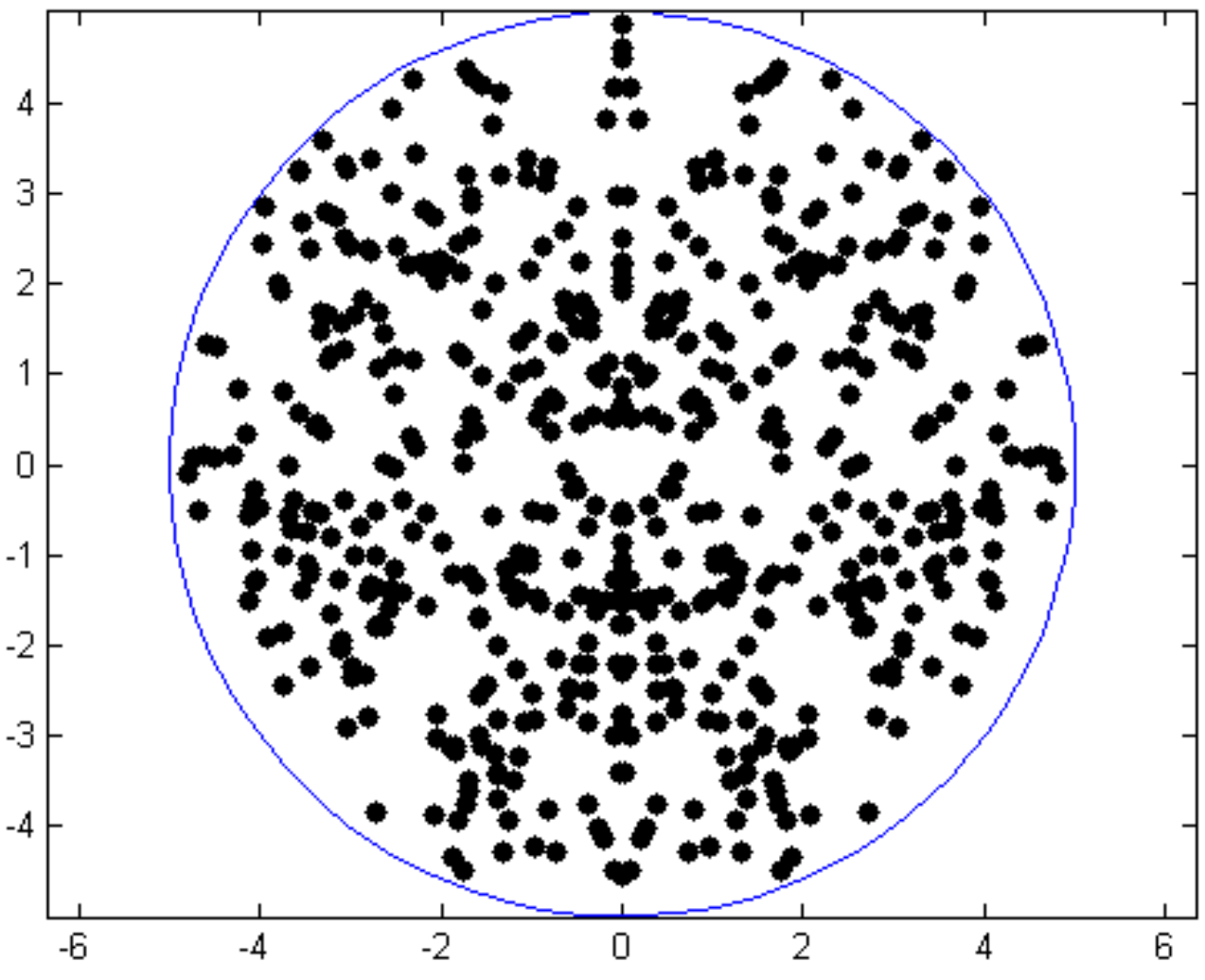

**Fig. 9.** The resulted pattern when T=45, N=20, S=20, D=1000.

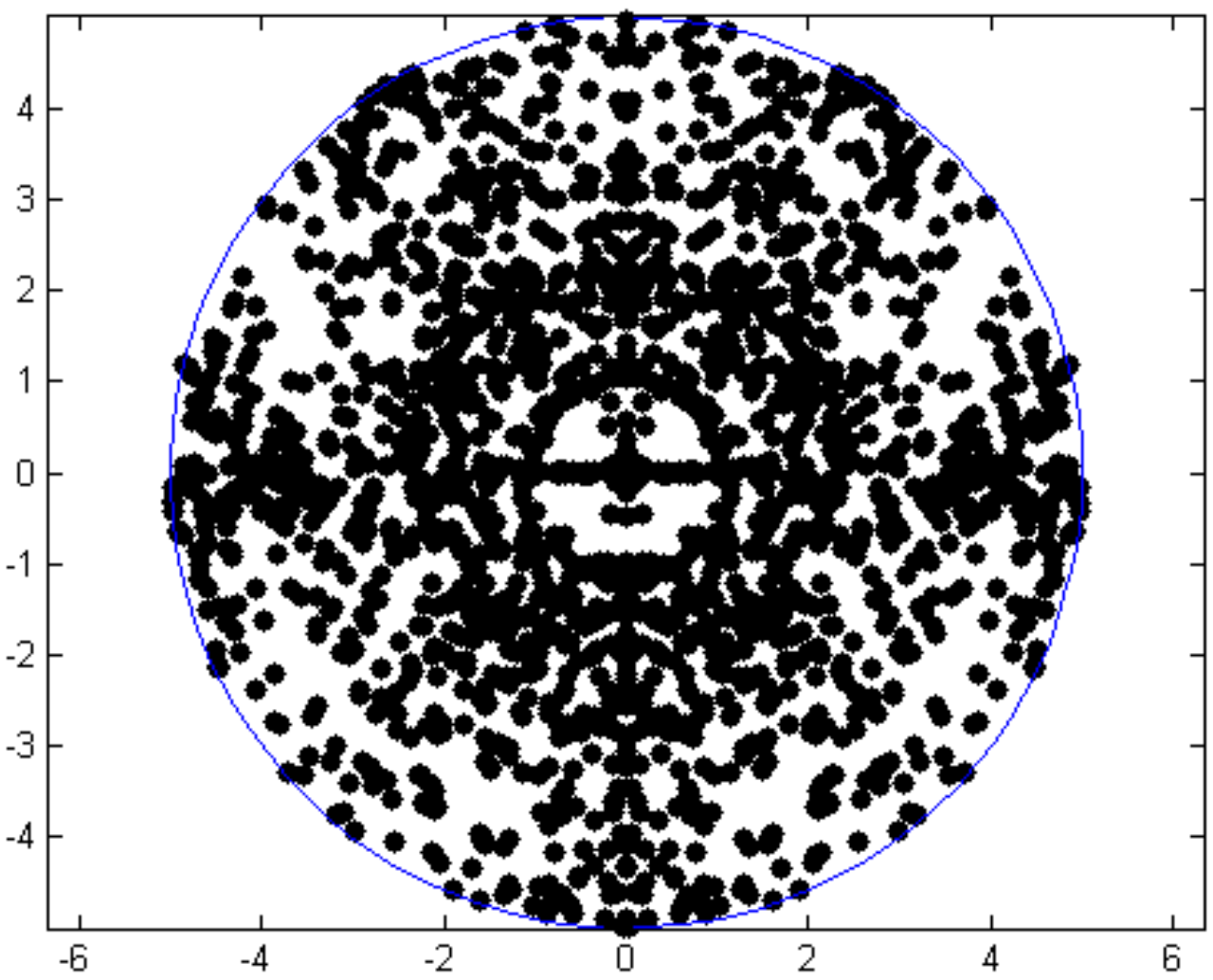

**Fig. 10.** The resulted pattern when T=30, N=20, S=20, D=400.

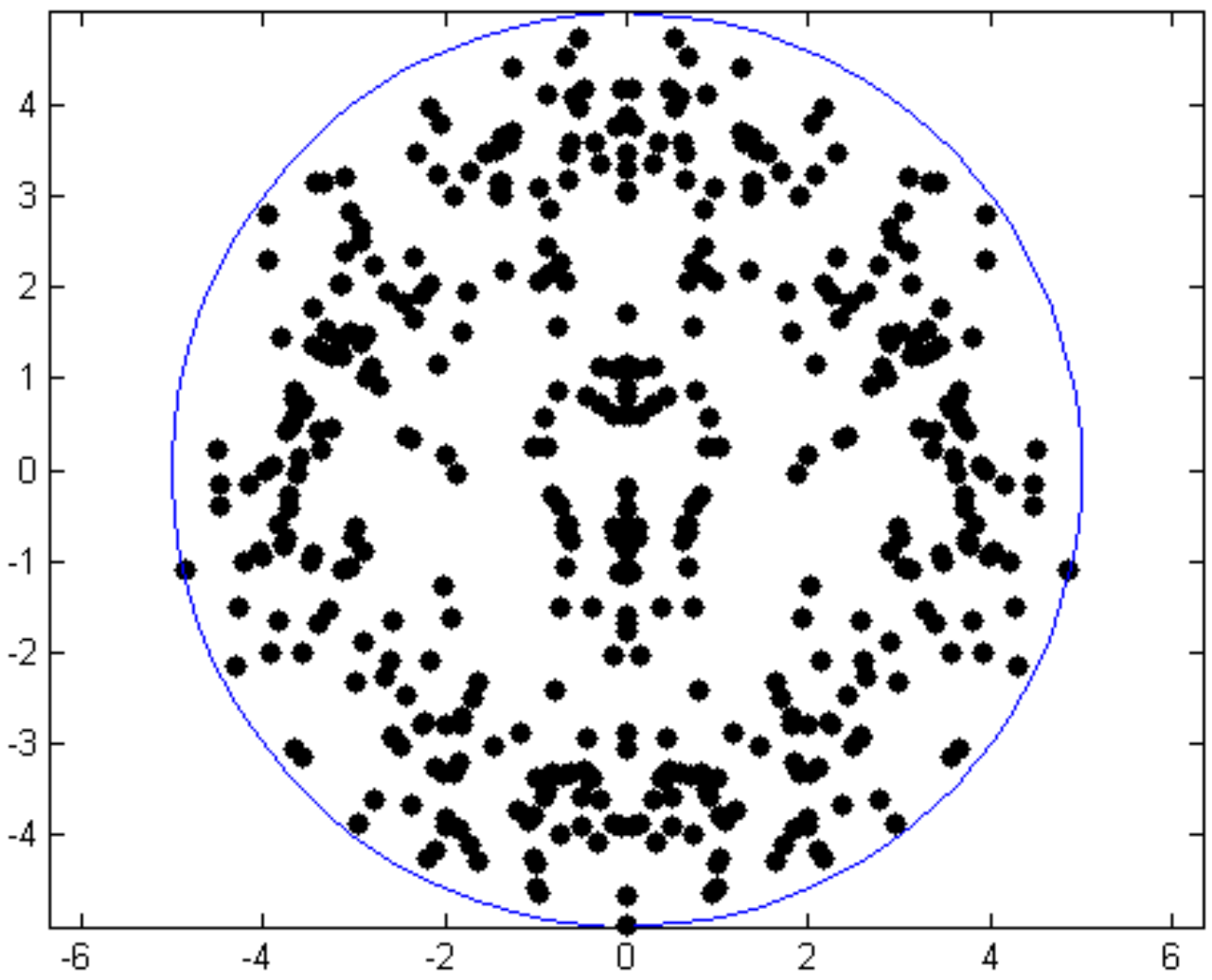

**Fig. 11.** The resulted pattern when T=60, N=20, S=20, D=1500.

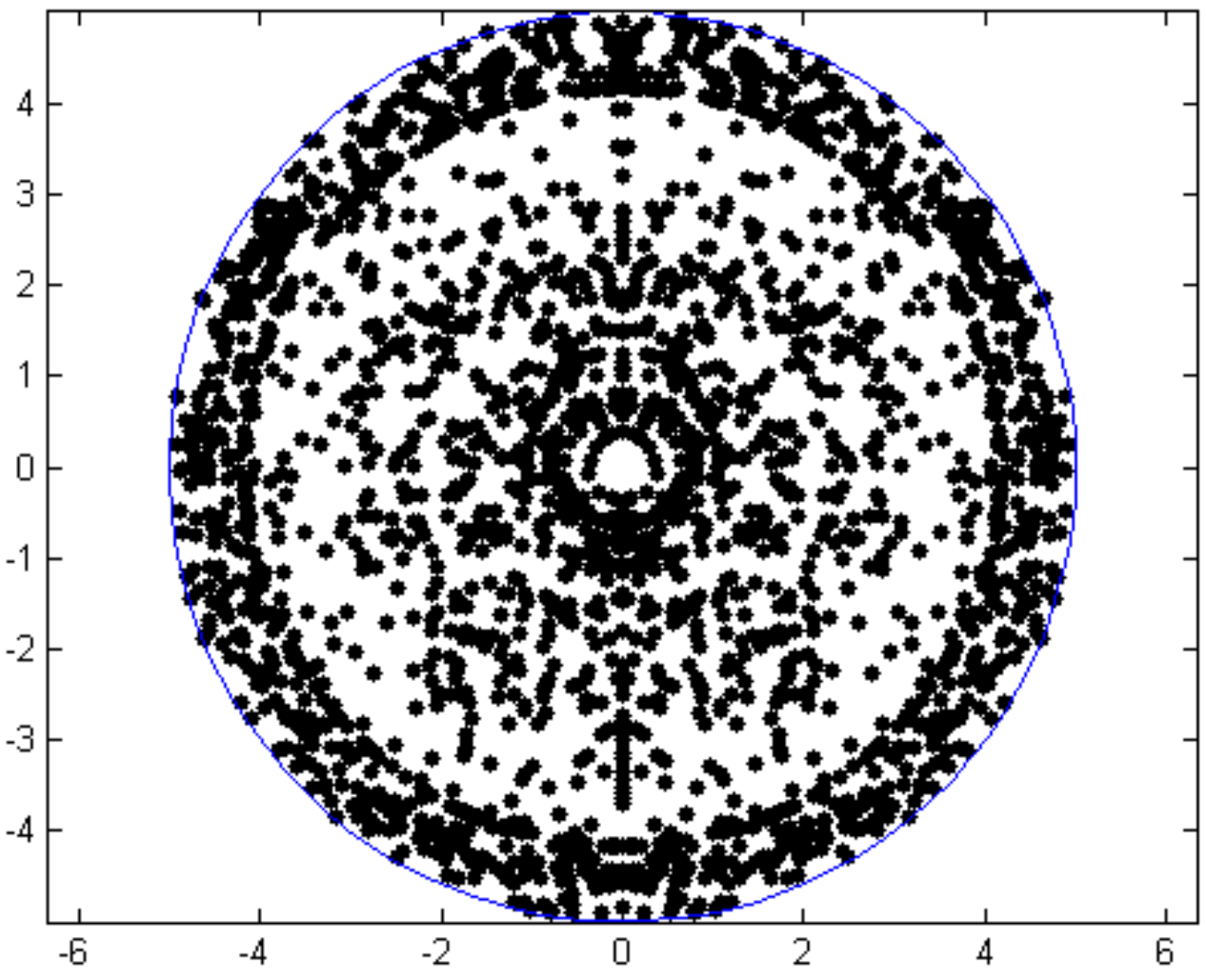

**Fig. 12.** The resulted pattern if T=120, N=20, S=15, D=1500.

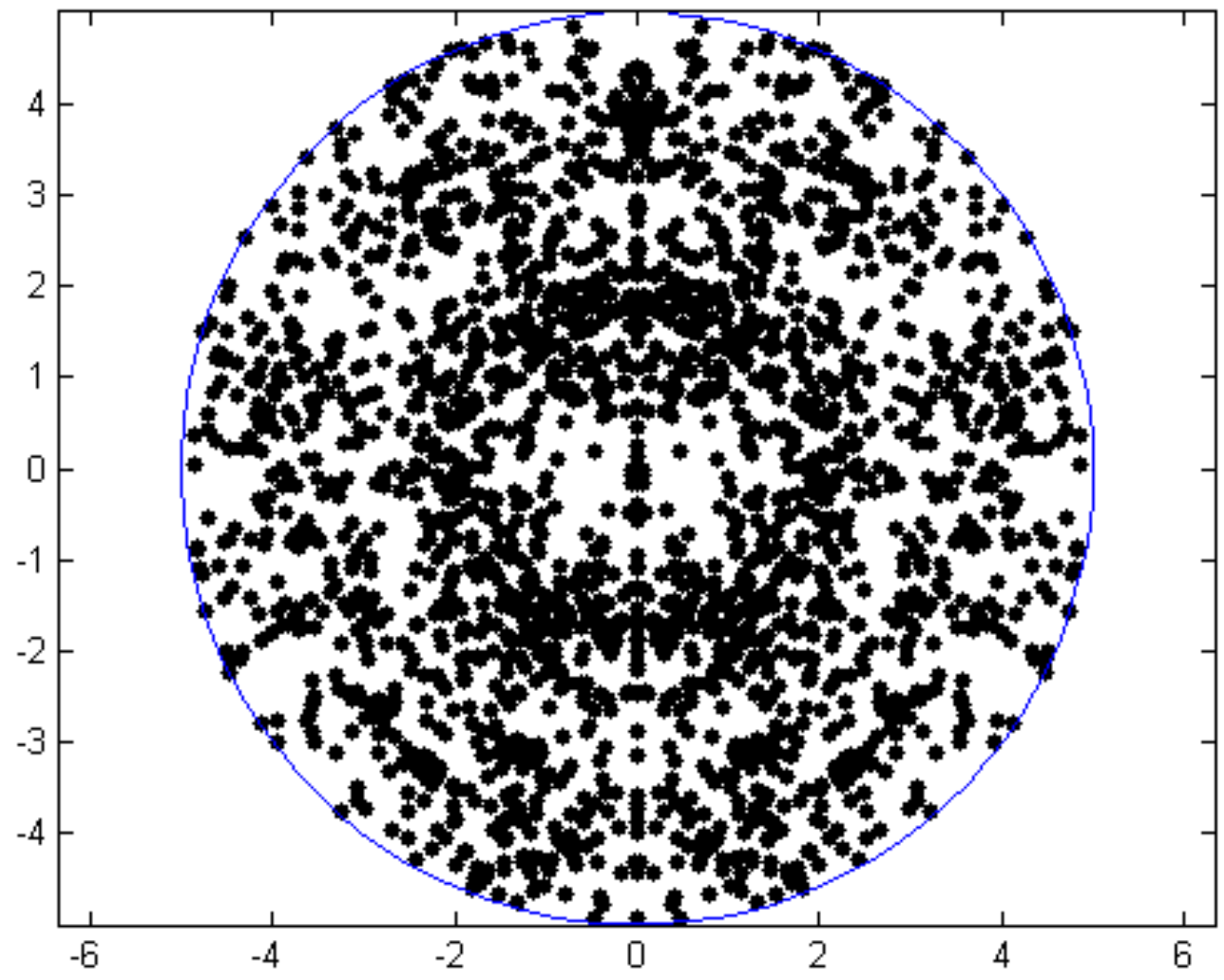

**Fig. 13.** The resulted pattern if T=120, N=20, S=15, D=1500.

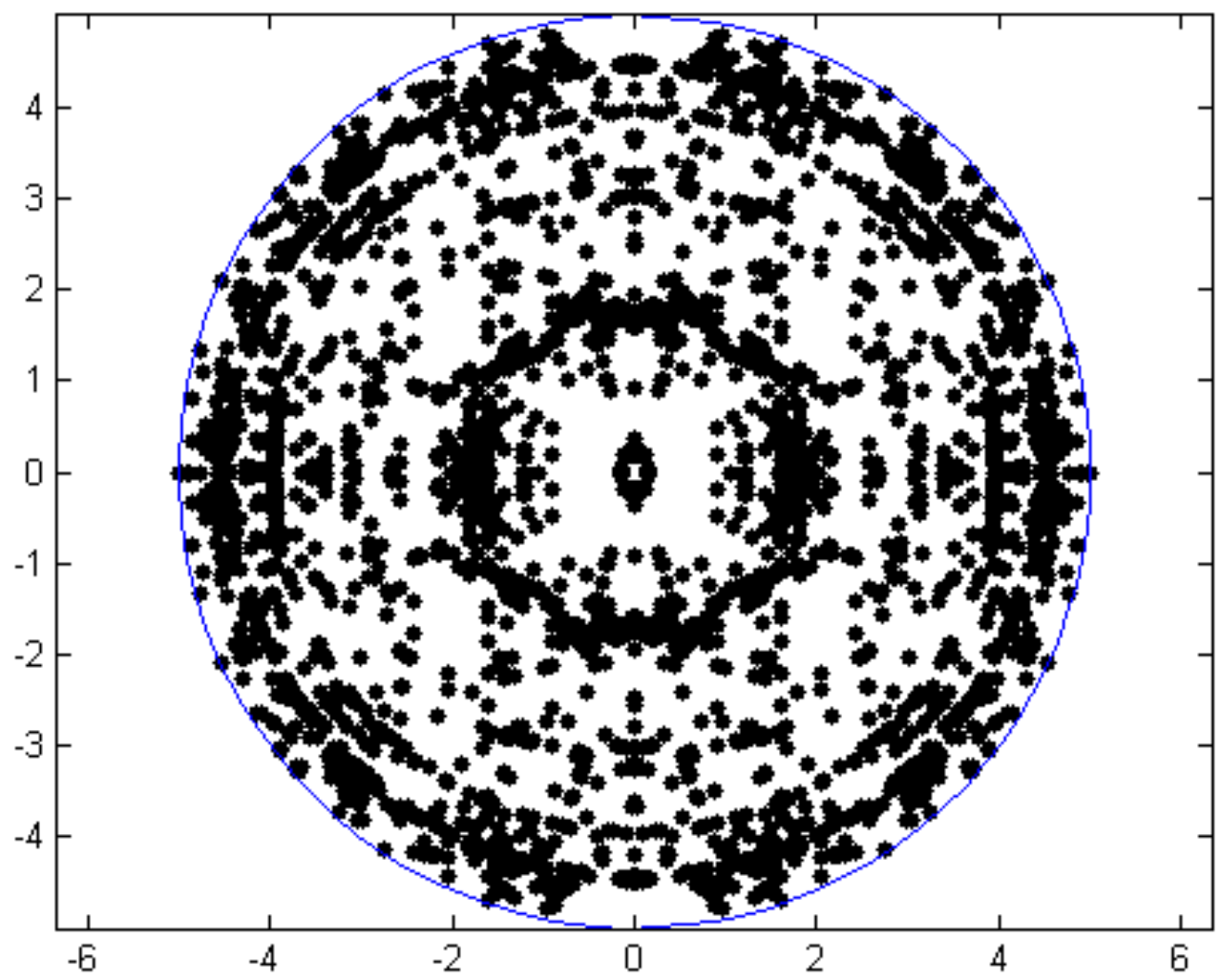

**Fig. 14.** The resulted pattern when T=15, N=20, S=10, D=50.

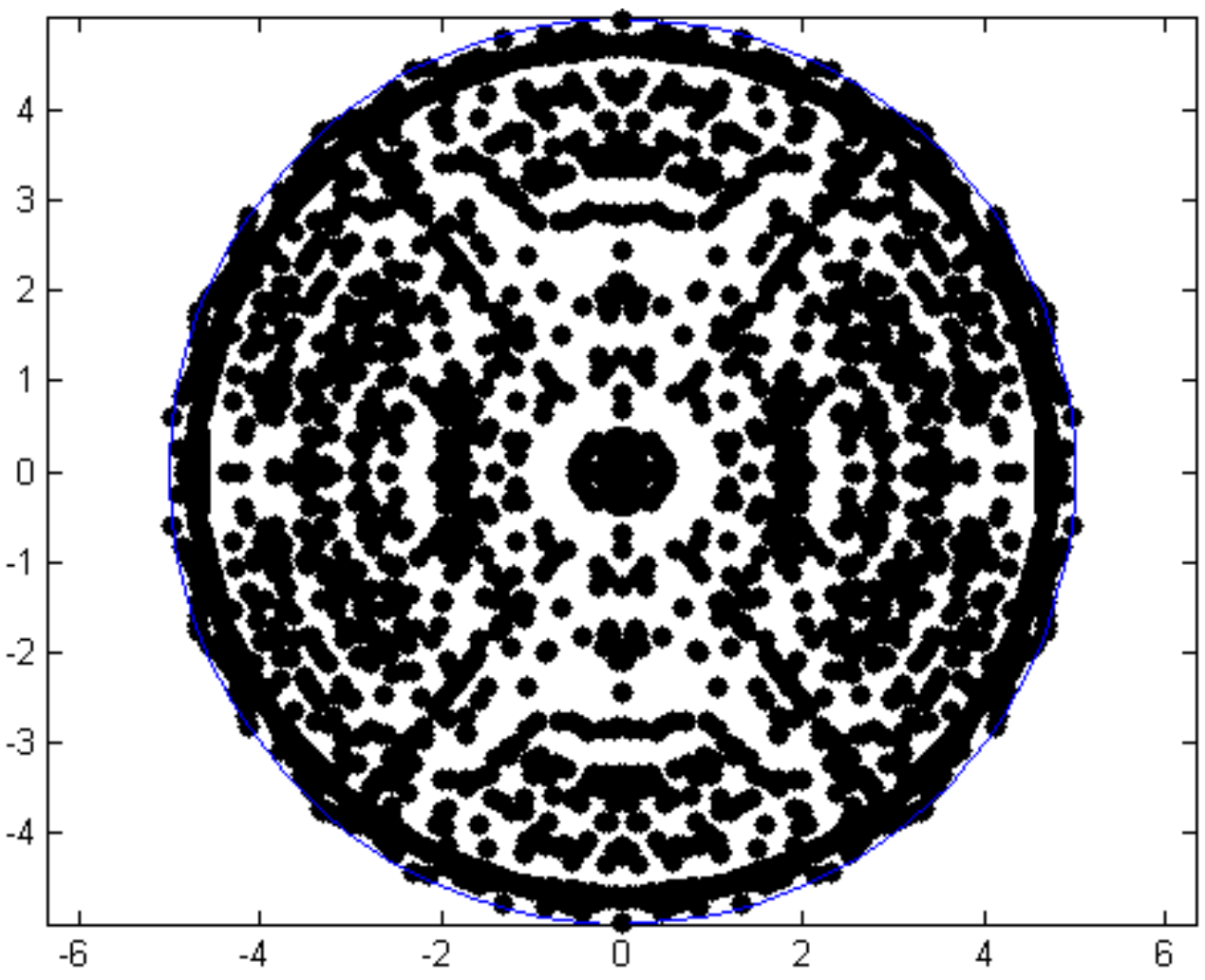

**Fig. 15.** The resulted pattern when T=15, N=20, S=20, D=50.

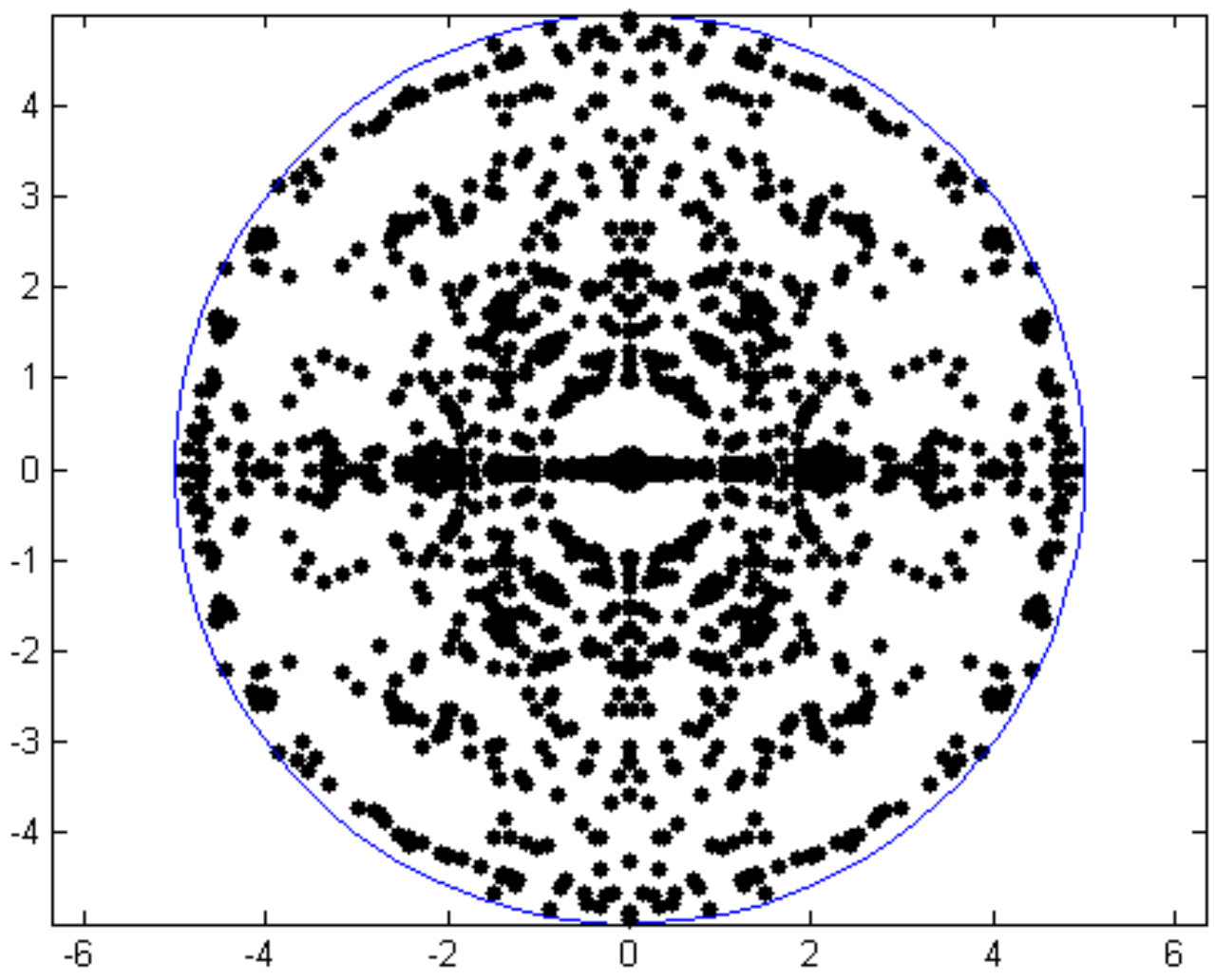

**Fig. 16.** The resulted pattern when T=30, N=20, S=15, D=100.

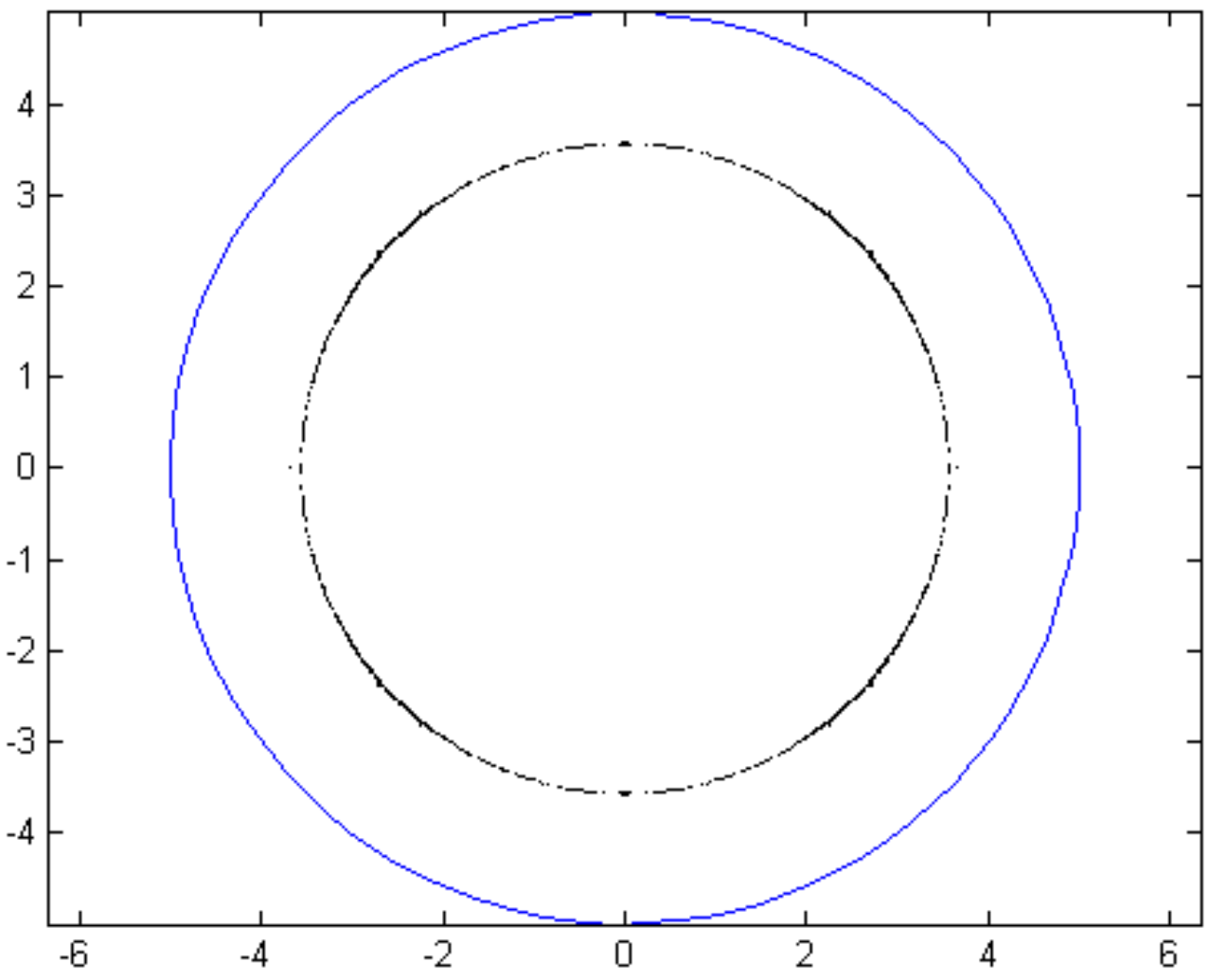

**Fig. 17.** The resulted pattern when T=120, N=1, S=1, D=100.

To provide a quantitative evaluation of the outcome of the simulations, we will investigate the distribution of the distances of the particles from the origin. The distribution will give deeper information about the spatial distribution of the particles. Fig. 18 to Fig. 20 show the distribution of the distances of the particles from the origin for patterns that have complex geometrical structure. Fig. 21 to Fig. 23 show the distribution of the distances of the particles from the origin for patterns that have face-like structure. Common features of the distributions within the same category can be seen. The number of symmetrical particles NS in the resulted patterns is also introduced to give a quantitative measure, as can be seen from the given examples (Fig.18 to Fig.23), complex geometrical structure patterns have far higher number of symmetrical particles than the face-like structure patterns. The value of NS is listed below each figure.

Entropy has limitations in representing the spatial arrangement of the pattern. To provide a better quantitative measure of the resulted patterns, compressibility (C) is considered. The compressibility metric is simply the ratio between the original and the compressed image using the JPEG method. Fig. 18 to Fig. 23 show how this measure deviates from random permutations of the same pattern, where Cr represents the compressibility when the positions of the particles were randomly moved. The results show a significant distance between the random and the original pattern.

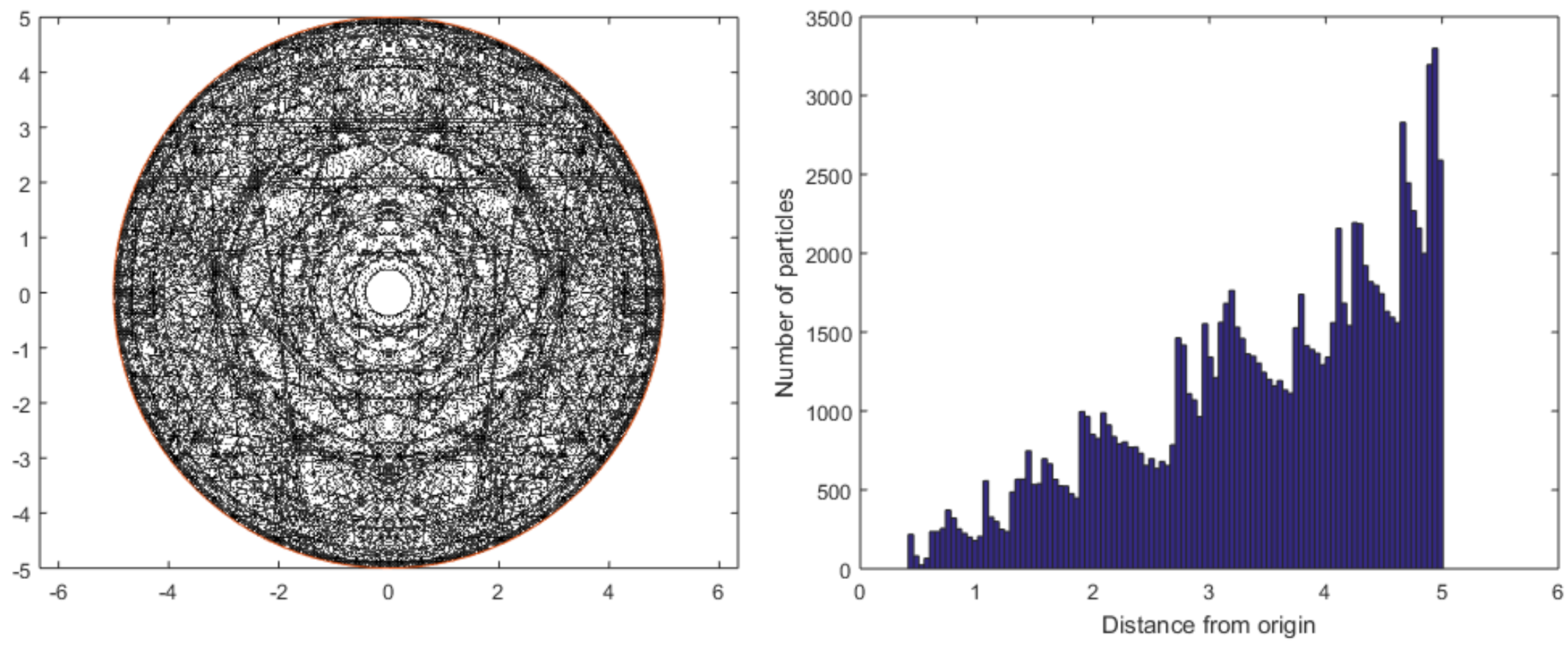

**Fig. 18.** The resulted pattern and distribution when T=120, N=20, S=1, D=75, NS=109702, C=2.0009, Cr=1.5948.

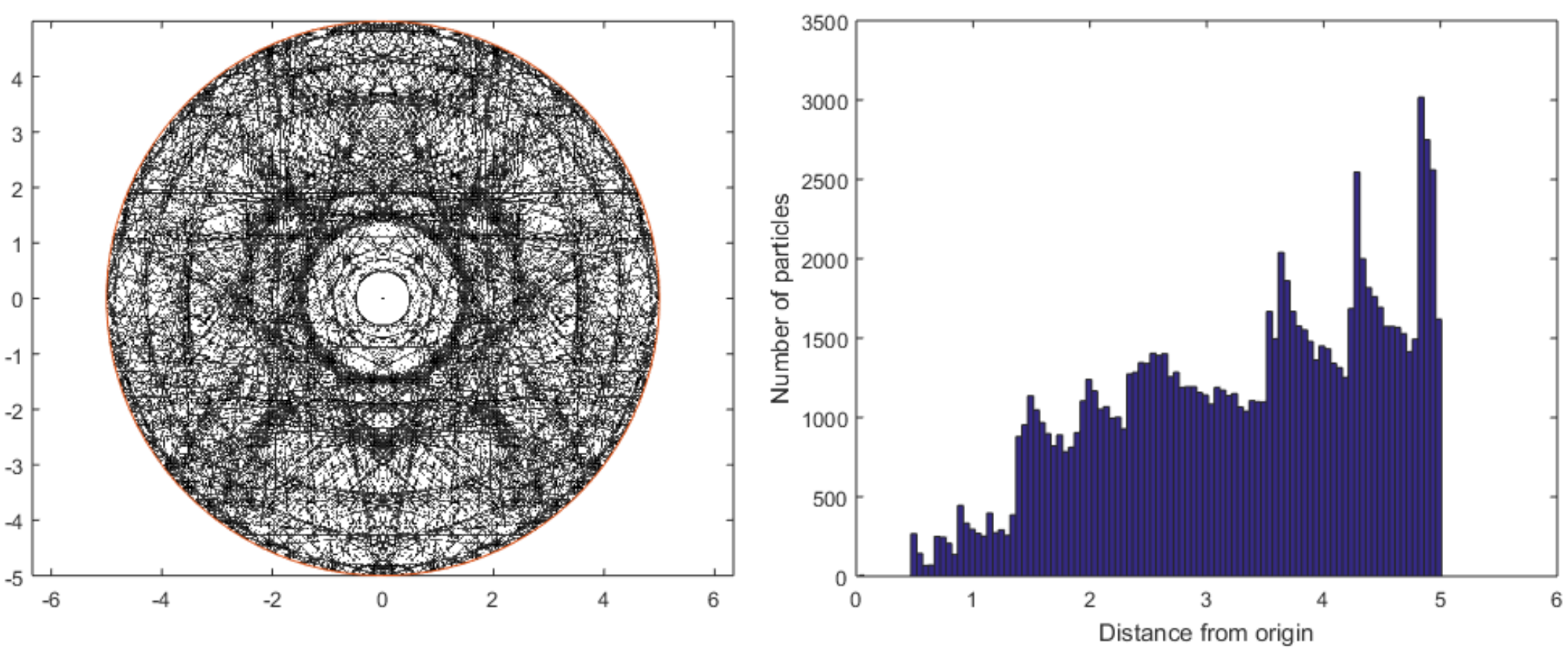

**Fig. 19.** The resulted pattern and distribution when T=120, N=20, S=1, D=75, NS=104364, C= 2.0046, Cr=1.5948.

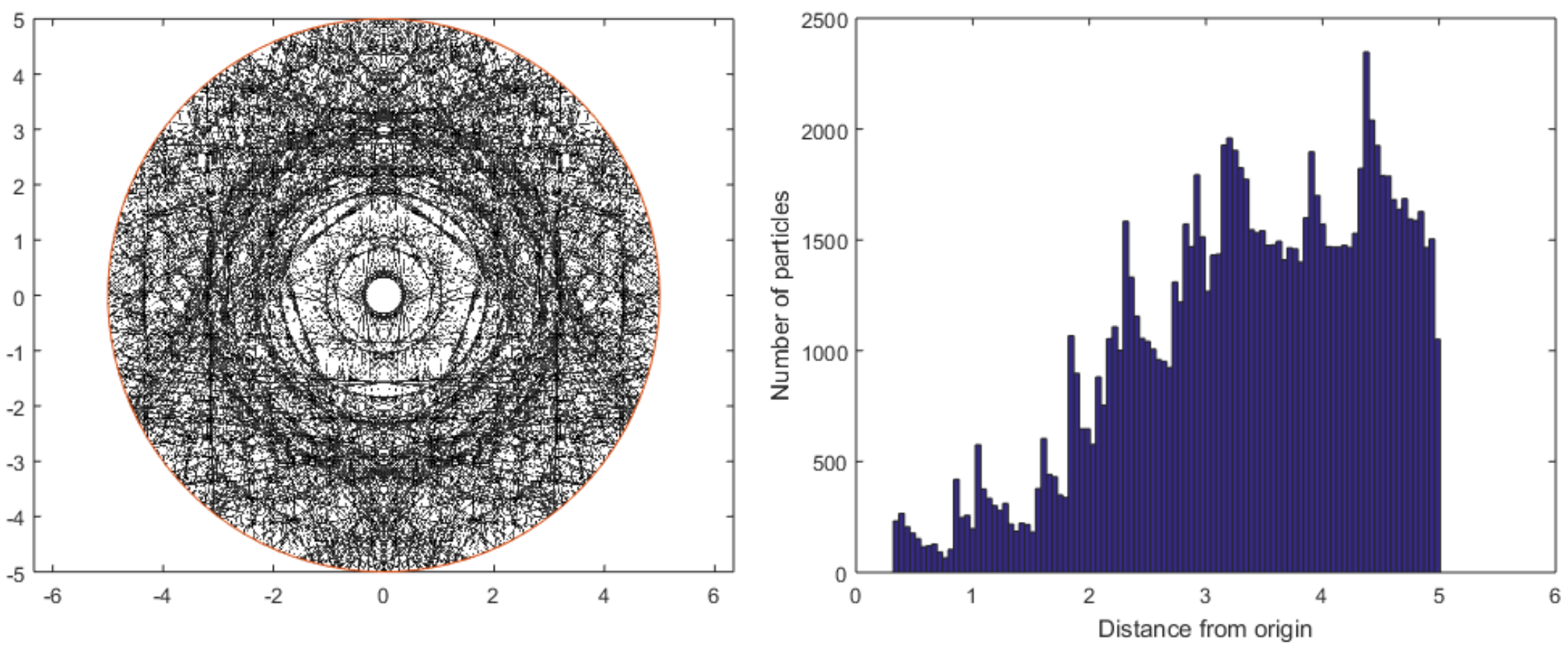

**Fig. 20.** The resulted pattern and distribution when T=120, N=20, S=1, D=75, NS=105596, C= 1.9972, Cr=1.5963.

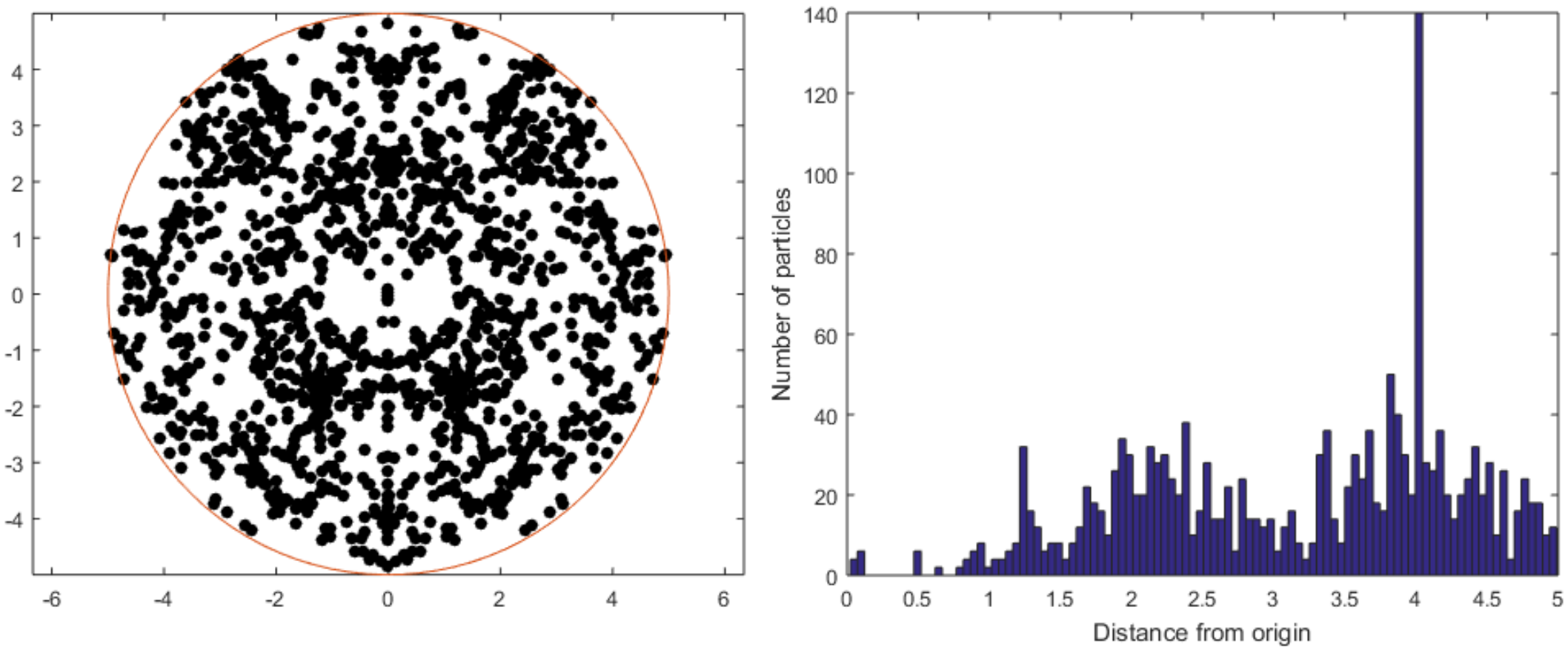

**Fig. 21.** The resulted pattern and distribution when T=120, N=20, S=20, D=400, NS=1678, C= 2.7277, Cr= 2.2373.

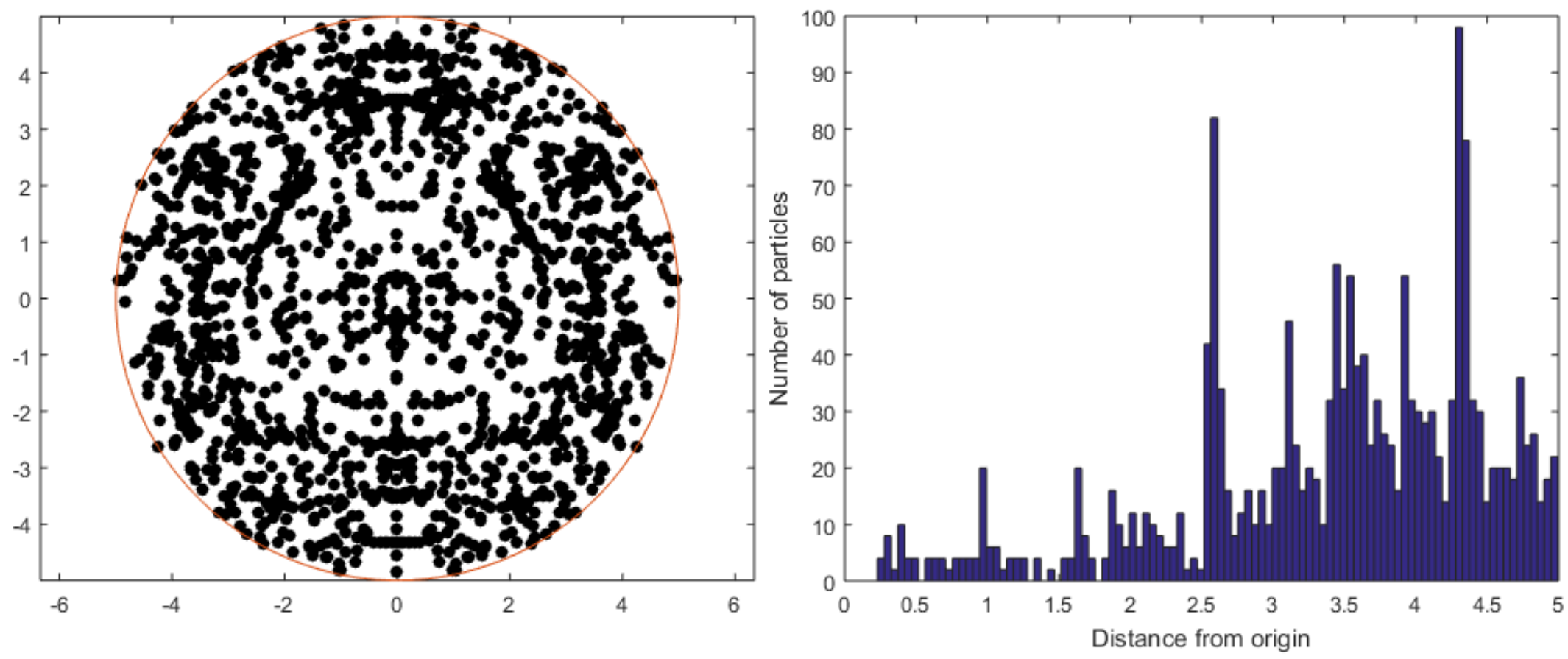

**Fig. 22.** The resulted pattern and distribution when T=120, N=20, S=20, D=400, NS=1778, C= 2.5748, Cr= 2.2657.

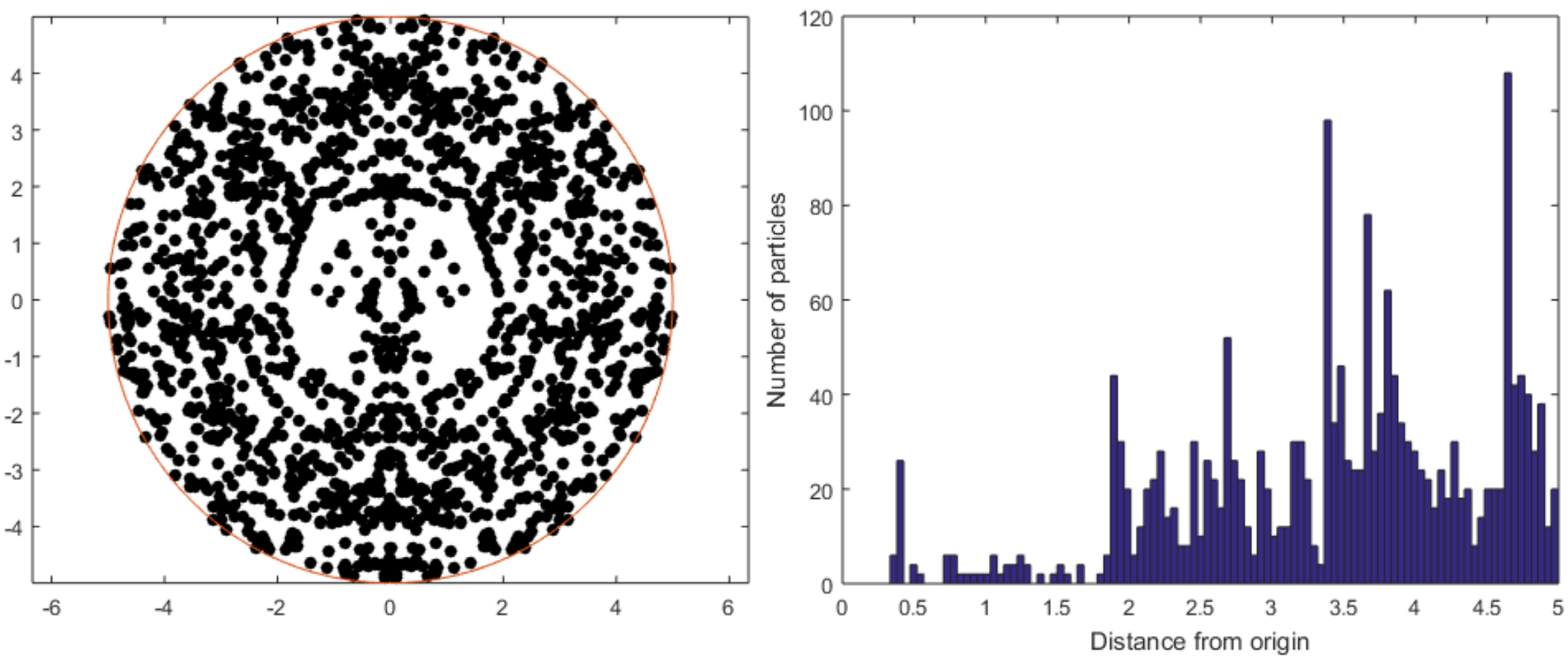

**Fig. 23.** The resulted pattern and distribution when T=120, N=20, S=20, D=400, NS=1912, C= 2.5644, Cr=2.2681.

To investigate how the initial conditions affect the resulted patterns, a one particle will be used, the direction of the particle will be fixed to $\pi/2$, the y coordinate will be fixed to zero, and the x coordinate will vary from 0.2r to 0.9r, where r is the radius of the circle. Fig. 24 to Fig. 28 show how the pattern changes by changing the initial position of the particle. The initial position of the particle limits the positions that the particle can visit, and that affects the complexity of the resulted pattern.

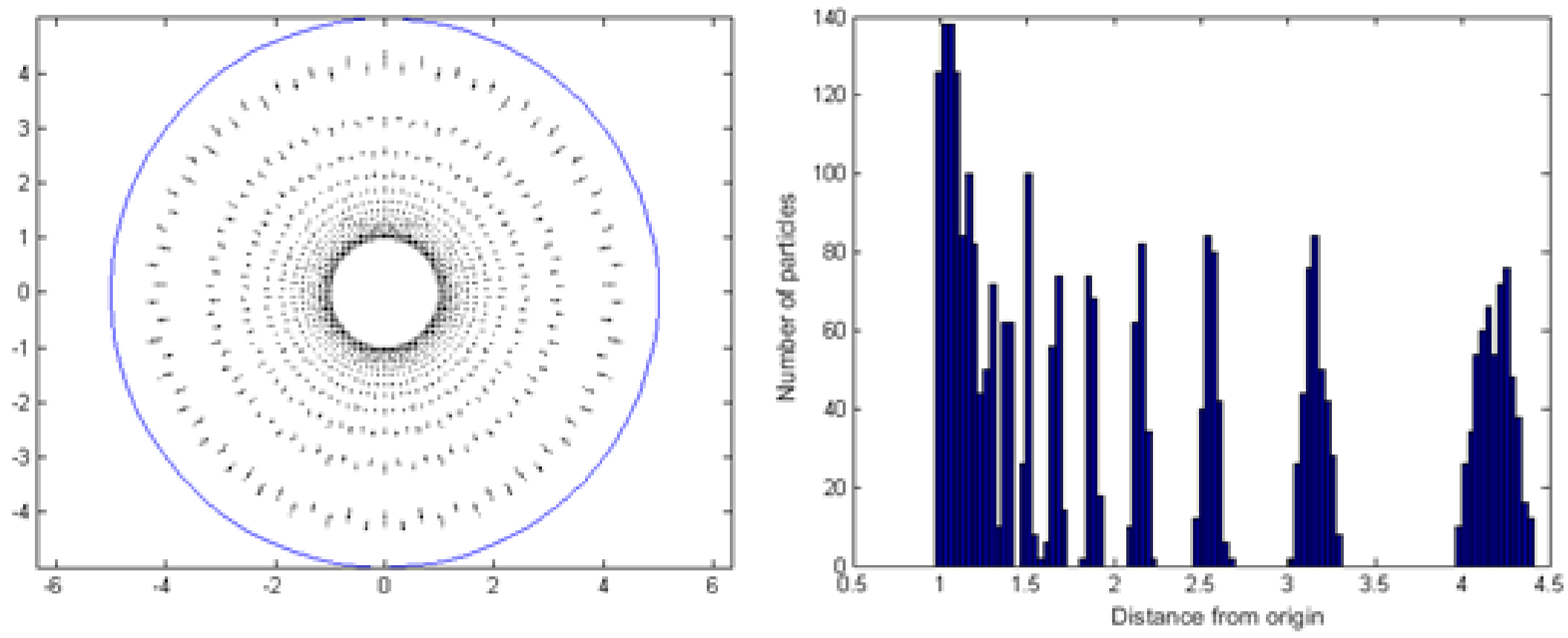

**Fig. 24.** The resulted pattern and distribution when T=300, N=1, S=1, D=25, X=0.2r.

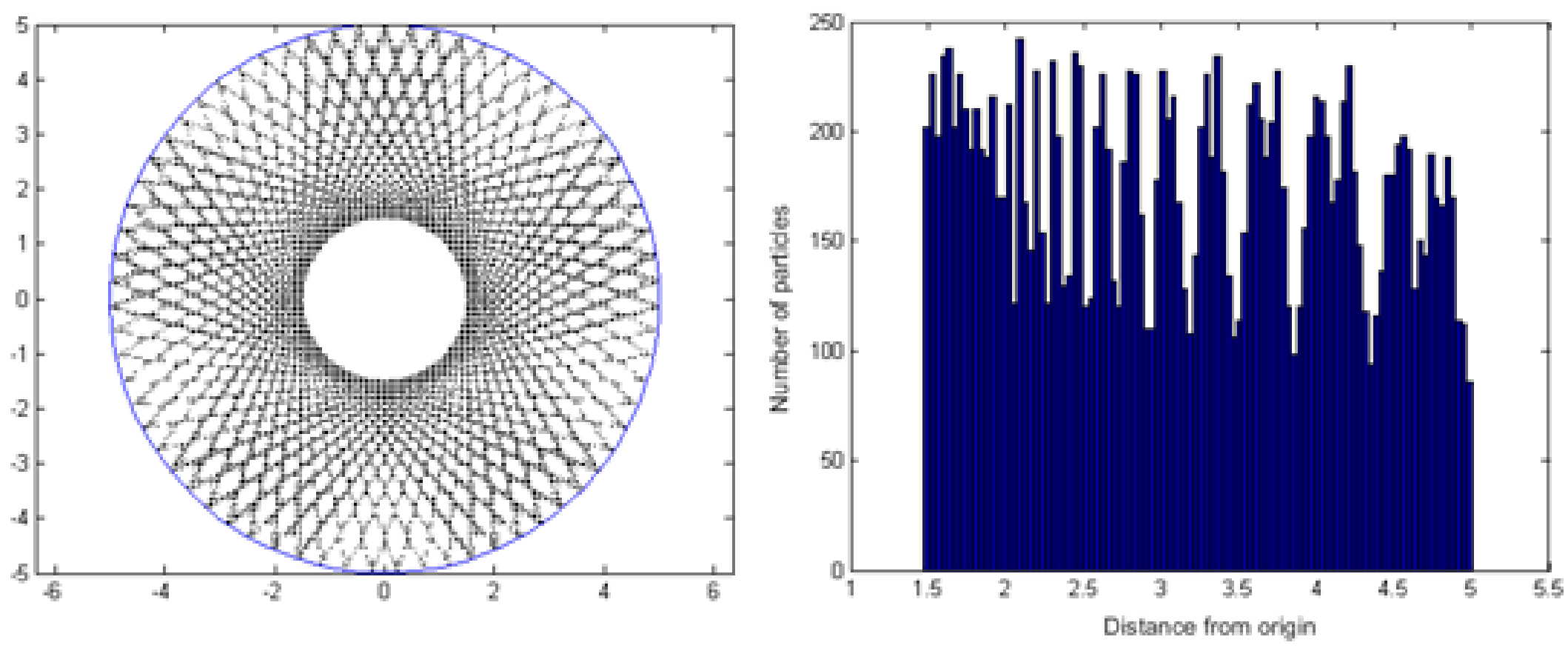

**Fig. 25.** The resulted pattern and distribution when T=300, N=1, S=1, D=25, X=0.3r.

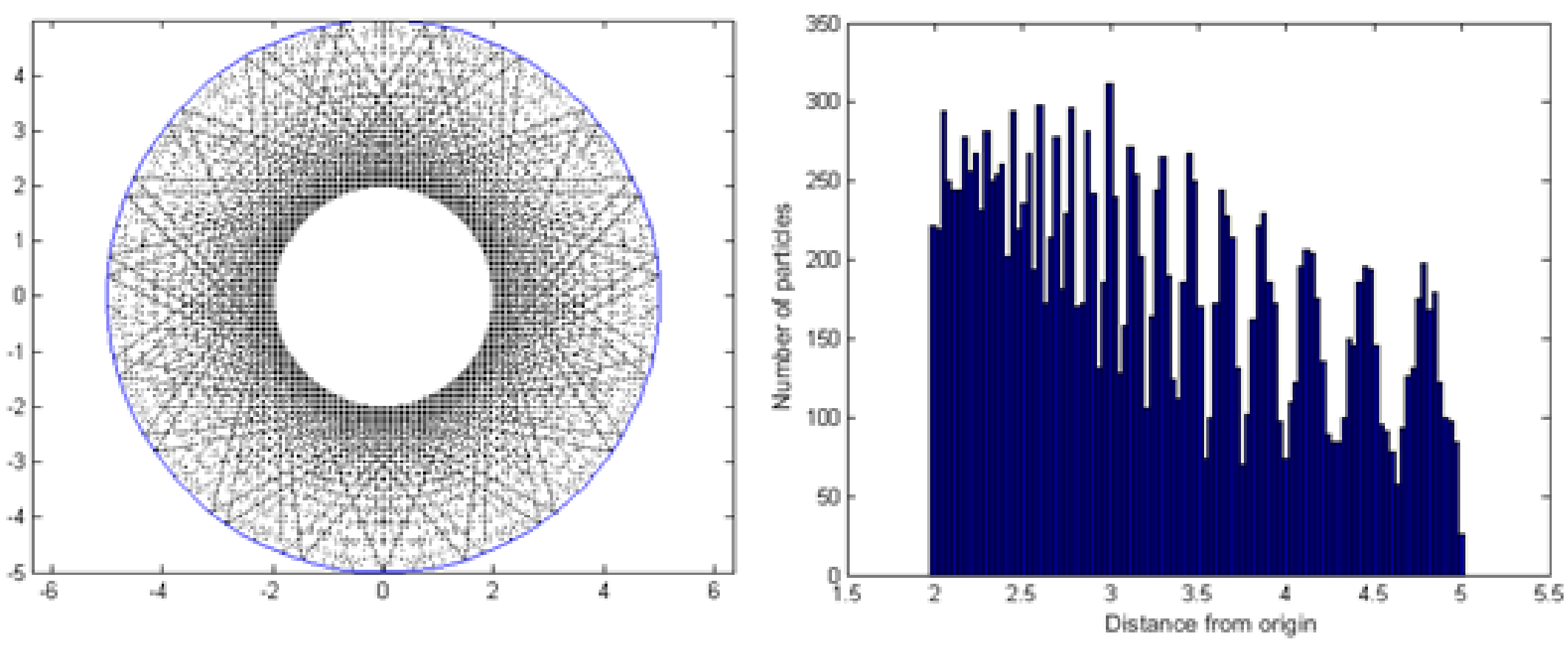

**Fig. 26.** The resulted pattern and distribution when T=300, N=1, S=1, D=25, X=0.4r.

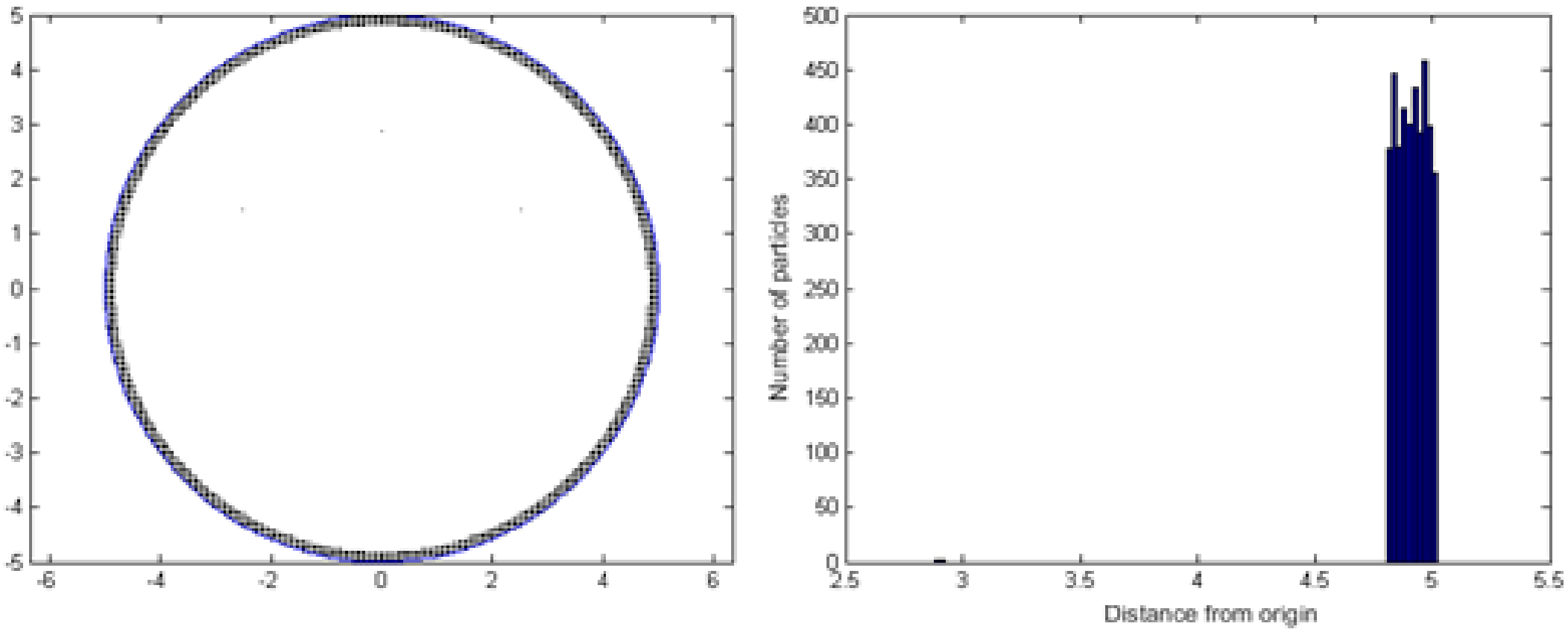

**Fig. 27.** The resulted pattern and distribution when T=300, N=1, S=1, D=25, X=0.5r.

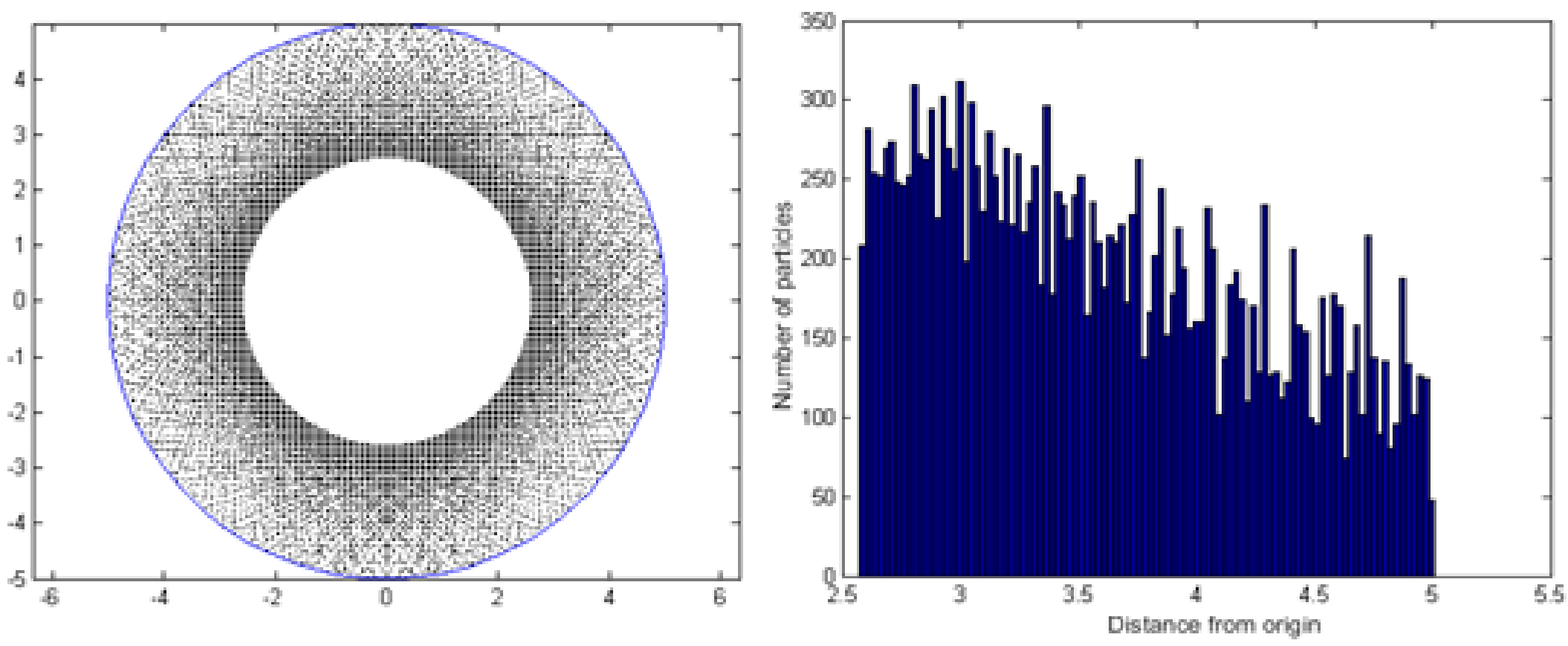

**Fig. 28.** The resulted pattern and distribution when T=300, N=1, S=1, D=25, X=0.8r.

### Discussion

The previous section showed how simple rules namely symmetry rules could produce very complex patterns. Two types of patterns were observed, the first one is the complex geometrical structure such as Fig. 20, and the second one is the face-like structure such as Fig. 13. We also showed how the initial conditions affect the resulted patterns. The distribution of the distances of the particles from the origin could help in determining the presence of a structure in the resulted pattern, it could also help in distinguishing between different patterns, for example, in the complex geometrical structure case, the number of particles at different distances has a periodic nature, i.e. it increases then decreases and then increases again and so on. Other patterns such as the circle has a very narrow distribution. The main difference between these two categories is the number of symmetrical particles where complex geometrical structure patterns have far higher number of symmetrical particles than the other category. The main difference between the complex geometrical structure and the face-like structure in term of input parameters is the parameter D (the number of network nodes per distance unit, where the higher the number of nodes, the lower the number of symmetrical particles in the resulted pattern). A lower number of nodes will produce complex geometrical structures while less number of nodes will produce patterns in the second category. This shows that different levels of granularity produce different emergent behaviours. However, the use of low level measures such as the distribution, the density of the particles or the number of symmetrical particles in describing the resulted patterns is rather limited and higher level approaches such as shape recognition and face recognition should be used.

The observed patterns showed how the proposed framework was able to model non-local and time-independent phenomena. In particular we showed how simple non-local and time-independent rules give rise to complex symmetrical patterns. This particular example was presented to show the effectiveness of the proposed framework. The broader aim of the proposed framework is to model more complex real-world phenomena, such as fluid dynamics where the macroscopic behaviour of the fluid is govern by rules governing the interactions of the molecules. Another example is the temperature where the rules governing the interactions of gas molecules give rise to emergent property such as the temperature. The proposed framework could also find applications in neural networks; for instance, the proposed framework could help in modelling non-local interactions between neurons that do not have direct connections. It could help in modelling how a hypothetical event in the future will affect the current state of the network. Finally, it could help in optimizing a global criterion over the entire neural network. Other examples include the quantum phenomena described earlier, however modelling these phenomena is beyond the scope of this paper. Future work will investigate the use of other quantitative measures to describe the resulted patterns, it will also investigate how the resulted patterns will be affected when we put some constraints on the distribution of the distances of the particles from the origin. Future work will also investigate the use of other rules particularly rules that may give rise to the complex behaviour we see in natural phenomena.

## Conclusion

This paper has presented a novel approach to produce emergent behaviour from simple underlying rules. The proposed approach goes beyond existing approaches by using a dynamic, non-local, and time independent approach that uses a network-like structure to implement the laws or the rules, where the mathematical equations representing the rules are converted to a series of switching decisions carried out by the network on the particles moving in the network. The results showed that the proposed approach was able to produce interesting symmetrical patterns.